\documentclass{aa}
\usepackage{graphicx}
\usepackage{txfonts}
\usepackage{natbib}

\begin{document}

   \title{FIRST-based survey of compact steep spectrum sources}

   \subtitle{V. Milliarcsecond-scale morphology of CSS objects} 

   \author{M. Kunert-Bajraszewska\inst{1}
          \and A. Marecki\inst{1}
          }

   \offprints{M. Kunert-Bajraszewska\\ \email{magda@astro.uni.torun.pl}}

   \institute{Toru\'n Centre for Astronomy, N. Copernicus University,
              87-100 Toru\'n, Poland
             }

   \date{Received 8 September 2006; Accepted 7 March 2007}

\abstract
{}
{Multifrequency VLBA observations of the final group of ten objects in a sample 
of FIRST-based compact steep spectrum (CSS) sources are presented. The sample
was selected to investigate whether objects
of this kind could be relics of radio$-$loud AGNs switched off at very early
stages of their evolution or possibly to indicate intermittent activity.}
{Initial observations were made using MERLIN at 5\,GHz. The sources have now
been observed with the VLBA at 1.7, 5 and 8.4\,GHz in a snapshot mode with
phase-referencing. The resulting maps are presented along with unpublished
8.4-GHz VLA images of five sources.} 
{Some of the sources discussed here show a complex radio morphology and
therefore a complicated past that, in some cases, might indicate intermittent
activity. One of the sources studied -- 1045+352 -- is known as a powerful
radio and infrared-luminous broad absorption line (BAL) quasar. 
It is a
young CSS object whose asymmetric two-sided morphology on a scale of several
hundred parsecs, extending in two different directions, may suggest intermittent
activity. The young age and compact structure of 1045+352 is consistent with
the evolution scenario of BAL quasars. It has also been confirmed that
the submillimetre flux of 1045+352 can be seriously contaminated by synchrotron 
emission.}
{} 


\maketitle


\section{Introduction}

Following early hypotheses \citep{pm82,c85} suggesting that the
gigahertz-peaked spectrum (GPS) and compact steep spectrum (CSS) could be
young objects, \citet{r96} proposed an evolutionary scheme unifying three
classes of radio-loud AGNs (RLAGNs): symmetric GPS objects -- CSOs (compact
symmetric objects); symmetric CSS objects -- MSOs (medium-sized symmetric objects) and
large symmetric objects (LSOs). In this scheme GPS/CSO sources with linear sizes
less than 1\,kpc\footnote{For consistency with earlier papers in this field,
the following cosmological parameters have been adopted throughout this
paper: $H_0$=100${\rm\,km\,s^{-1}\,Mpc^{-1}}$ and $q_0$=0.5. Throughout this
paper, the spectral index is defined such that $S\propto\nu^{\alpha}$.}
would evolve into CSS/MSOs with subgalactic sizes ($<$20\,kpc) and these in
turn would eventually become LSOs during their lifetimes.
Two pieces of evidence definitely point towards GPS/CSS sources being young objects: 
lobe proper motions (up to 0.3c) giving kinematic ages as low as
$\sim$$10^3$~years for CSOs \citep{ocp98,gir03,pol03} and
radiative ages typically $\sim$$10^5$~years for MSOs \citep{mur99}.
Although these AGNs are small-scale objects, in some cases CSO/GPS sources are
associated with much larger radio structures that extend out to many kiloparsecs.
In these cases, it has been suggested that the CSO/GPS stage represents a period of 
renewed activity in the life cycle of the AGN \citep[and references
therein]{stan05}.
\citet{rb97} have also proposed a model in which extragalactic radio sources are 
intermittent on timescales of $\sim$$10^4$--$10^5$~years.
Following the above scenarios and also an earlier suggestion by \citet{r94}
and \citet{odea97} that there exists a large population of compact, 
short-lived objects, \citet{mar46, mks06} concluded that
the evolutionary track proposed by \citet{r96} is only one of many possible tracks.
A lack of stable fuelling from the black hole can inhibit the growth of a
radio source, and consequently it will never reach the LSO stage, at least in a given
phase of its activity. 

Observational support for the above ideas has been provided by
\citet{gug05}. They calculated the kinematic ages for a
sample of CSOs with well-identified hotspots. It appears that the kinematic
age distribution drops sharply above $\sim$500 years, suggesting that in many
CSOs activity may cease early. It is, therefore, possible that only some of them
evolve any further.
Our observations have shown that young, fading compact sources do indeed
exist \citep[][hereafter Papers II, III, and IV, respectively]{kmts05, mks06,
kun06}. A double source, 0809+404, described in Paper~IV is our best example
of a very compact -- i.e. very young -- fader. The VLBA multifrequency
observations have shown it to have a diffuse, amorphous structure, devoid 
of a dominant core and hotspots. \citet{gir05} have analysed the properties
of a sample of small-size sources and found a very good example of a
kiloparsec-scale fader (1855+37). It is to be noted that re-ignition of 
activity in compact radio sources is not ruled out.
In this paper -- the fifth and the last of the series -- VLBA observations
of 10 CSS and CSO sources that are potential candidates for compact faders or objects
with intermittent activity are presented. One of these sources, 1045+352,
is of particular interest not only because it has a puzzling radio
structure, but it also appears to be a broad absorption line (BAL) quasar.

As their name somewhat suggests, BAL quasars have very broad, blue-shifted absorption
lines arising from high-ionization transitions such as C\,IV, Si\,IV, N\,V, etc. (e.g
C\,IV 1549$\AA$). They constitute $\sim$10\% of the optically selected radio-quiet
quasars with the absorption arising from gas outflow at velocities up to $\sim$0.2\,c \citep{hewett03}.
In fact, BAL quasars have been divided into two categories, as 10\% of them also
show absorption troughs in low-ionization lines such as Mg\,II 2800$\AA$.
This group has been designated as LoBAL quasars and the others as HiBAL ones. 
The high ionization level and continuous absorption over a wide velocity
range is hard to reconcile with absorption by individual clouds. Rather, they 
indicate that BAL regions exist in both BAL and non-BAL quasars and evidence, accumulated
from optically selected BAL quasars, indicates an orientation hypothesis to explain
their nature. It would appear that BAL quasars are normal quasars seen along a particular
line of sight, e.g. a line of sight skimming the edge of the accretion disk
or torus \citep{weymann91, elvis00}. \citet{murray95} have proposed a model in which
the line of sight to a BAL quasar intersects an outflow or wind
that is not entirely radial, e.g. an outflow that initially emerges
perpendicular to the accretion disk and is then accelerated radially.

For quite a long time it was believed that BAL quasars were never
radio-loud. This view was challenged by \citet{becker97}, who discovered the first radio-loud
BAL quasar when using the VLA FIRST survey to select quasar candidates. Five 
radio-loud BAL quasars were then identified in NVSS by \citet{broth98}. Since then, 
the number of radio-loud BAL\,QSOs has increased considerably \citep{becker00,
menou01},
following identification of new quasar candidates selected from the FIRST survey.
Most of the BAL quasars in the \citet{becker00} sample tended to be compact at radio 
frequencies with either a flat or steep radio spectrum. Those with steep spectra could
be related to GPS and CSS sources. A variety of their spectral indices also suggested
a wide range of orientations, contrary to the interpretation favoured from optically 
selected quasars. Moreover, \citet{becker00} indicated that the frequency of BAL
quasars in their sample was significantly greater (factor $\sim$2) than
inferred from optically selected samples and that the frequency of BAL
quasars appeared to show a complex dependence on radio loudness.

The radio morphology of BAL quasars is important because it can
indicate inclination in BALs, and therefore yields a direct 
test of the orientation model. However, information about the radio
structure of BAL quasars is still very limited. Prior to 2006, only
three BAL quasars, FIRST J101614.3+520916 \citep{gregg00}, PKS 1004+13
\citep{wills99},
and LBQS 1138$-$0126 \citep{broth02} were known to have a
double-lobed FR\,II radio morphology on kiloparsec scales, although this
interpretation was doubtful for PKS 1004+13 \citep{gopal00}. 
Recently, the population of FR\,II-BAL quasars has increased to ten
objects (excluding PKS 1004+13) following the discoveries of \citet{gregg06} and
\citet{zhou06}, although some of these still require confirmation.  
Their symmetric structures indicate an ``edge-on'' orientation, which in
turn supports an alternative hypothesis described as ``unification by time'',
with BAL quasars characterised as young or recently refuelled quasars
\citep{becker00, gregg00}. There has been only one attempt (at 1.6\,GHz
with the EVN) to image radio
structures of the smallest (and possibly the youngest) BAL quasars
\citep{jiang03} from the \citet{becker00} sample.
This paper presents high frequency VLBA images of another very
compact BAL quasar --- 1045+352, which makes it the BAL quasar with the best
known radio structure to date.

\begin{table*}[t]
\caption[]{Basic parameters of target sources}
\begin{center}
\begin{tabular}{@{}c c c c l l c c c c r c@{}}
\hline
\hline
~~~Source & RA & Dec & ID&
\multicolumn{1}{c}{$m_{R}$}&
\multicolumn{1}{c}{\it z}&
\multicolumn{1}{c}{$S_{1.4\,GHz}$}&
\multicolumn{1}{c}{log$P_{1.4\mathrm{GHz}}$}&
\multicolumn{1}{c}{$S_{4.85\,GHz}$}&
\multicolumn{1}{c}{$\alpha_{1.4\mathrm{GHz}}^{4.85\mathrm{GHz}}$}&
\multicolumn{1}{c}{LAS}&
\multicolumn{1}{c}{LLS}\\
~~~Name   & h~m~s & $\degr$~$\arcmin$~$\arcsec$ &  & & &
\multicolumn{1}{c}{mJy}&
\multicolumn{1}{c}{$\rm W~Hz^{-1}$} & 
\multicolumn{1}{c}{mJy}& &

\multicolumn{1}{c}{$\arcsec$} &$h^{-1}~{\rm kpc}~~~$ \\
~~~(1)& (2)& (3) &(4)&
\multicolumn{1}{c}{(5)}&
\multicolumn{1}{c}{(6)}&
\multicolumn{1}{c}{(7)}&
\multicolumn{1}{c}{(8)}&
\multicolumn{1}{c}{(9)}&
\multicolumn{1}{c}{(10)}&
\multicolumn{1}{c}{(11)}&
\multicolumn{1}{c}{(12)}\\
\hline
~~~1045+352 & 10 48 34.247 & 34 57 24.99  &Q  &20.86&1.604      &1051&27.65&439&$-$0.70
&$\sim$0.50 &2.1~~~\\
~~~1049+384 & 10 52 11.797 & 38 11 43.83  &G  &20.76&1.018      &712 &27.04&205&$-$1.00
&0.14  &0.6~~~\\
~~~1056+316 & 10 59 43.236 & 31 24 20.59  &G  &21.10&0.307$\ast$&459 &25.72&209&$-$0.63
&0.50 &1.4~~~\\
~~~1059+351 & 11 02 08.686 & 34 55 10.74  &G  &19.50&0.594$\ast$&702 &26.52&252&$-$0.82
&3.03 &11.5~~~\\
~~~1126+293 & 11 29 21.738 & 29 05 06.40  &EF  &---&---&729 &---&213&$-$0.99
&0.79 &---~~~\\
~~~1132+374 & 11 35 05.927 & 37 08 40.80  &G  &---&2.880      &638 &28.00&218&$-$0.86
&$\sim$0.30 &1.1~~~\\
~~~1302+356 & 13 04 34.477 & 35 23 33.93  &EF  &---&---&483 &---&185&$-$0.77
&$\sim$0.20 &---~~~\\
~~~1407+369 & 14 09 09.528 & 36 42 08.06  &q  &21.51&0.996$\ast$&538 &26.89&216&$-$0.73
&$\sim$0.25 &1.1~~~\\
~~~1425+287 & 14 27 40.281 & 28 33 25.78  &EF  &---&---&859 &---&198&$-$1.18
&0.75 &---~~~\\
~~~1627+289 & 16 29 12.290 & 28 51 34.25  &EF  &---&---&526 &---&162&$-$0.95
&$\sim$0.65 &---~~~\\
\hline
\end{tabular}
\end{center}

\vspace{0.5cm}
{\small
Description of the columns:
(1) source name in the IAU format;
(2) source right ascension (J2000) extracted from FIRST;
(3) source declination (J2000) extracted from FIRST;
(4) optical identification: G - galaxy, Q - quasar, EF - empty field, q - star-like object,
i.e. unconfirmed QSO;
(5) red magnitude extracted from SDSS/DR5;
(6) redshift;
(7) total flux density at 1.4\,GHz extracted from FIRST;
(8) log of the radio luminosity at 1.4\,GHz;
(9) total flux density at 4.85\,GHz extracted from GB6;
(10) spectral index between
1.4 and 4.85\,GHz calculated using flux densities in columns (7) and (9);
(11) largest angular size (LAS) measured in the
5-GHz MERLIN image -- in most cases, as a separation between the outermost
component peaks, otherwise ``$\sim$'' means measured in the image contour
plot;
(12) largest linear size (LLS).\\
$\ast$ photometric redshift extracted from SDSS/DR5
}

\label{table1}
\end{table*}

\section{The observations and data reduction}

The five papers of this series are concerned with a sample of 60~candidates
selected from the VLA FIRST catalogue \citep{wbhg97}\footnote{Official website:
http://sundog.stsci.edu} which could be weak CSS sources.
The sample selection criteria have been given in \citet{kun02} (hereafter Paper~I). All
the sources were initially observed with MERLIN at
5\,GHz and the results of these observations led to the selection of several
groups of objects for further study with MERLIN and the VLA (Paper~II), as well
as the VLBA and the EVN (Papers~III and~IV). The last of those groups contains
10~sources that, because of their structures (very faint ``haloes'' or possible 
core-jet structures), were not included in the other groups as they were less likely 
to be candidates for faders.
However, to complete the investigation of the primary sample, 1.7, 5, and 
8.4-GHz VLBA observations of 10~sources listed in Table~\ref{table1}
together with their basic properties,
were carried out on 13 November 2004 in a snapshot mode with
phase-referencing.\footnote{Including this
paper, the results of the observations of 46 sources out of 60 candidates from 
the primary sample have been published. The observations of 14 objects failed
for different reasons.} Each target source
scan was interleaved with a scan on a phase reference source
and the total cycle time (target and phase-reference) was
$\sim$9~minutes including telescope drive times, with $\sim$7\,minutes actually
on the target source per cycle. The cycles for a given target-calibrator
pair were grouped and
rotated round the three frequencies, although the source 1059+351 was
only observed at 1.7\,GHz with the VLBA because of its very low flux
density as measured at 5\,GHz by MERLIN (13 mJy).
 
The whole data reduction process was carried out using standard AIPS
procedures but, in addition to this, corrections for Earth orientation parameter 
(EOP) errors introduced by the VLBA correlator also had to be made. 
For each target source and at each frequency, the corresponding
phase-reference source was mapped, and the phase errors so determined were
applied to the target sources, which were then mapped using a few cycles of
phase self-calibration and imaging.
For some of the sources a final amplitude self-calibration was also applied.
IMAGR was used to produce the final ``naturally weighted'', total intensity images
shown in Figs.~\ref{1045+352_maps} to~\ref{1627+289_maps}.
Three of the ten sources (1056+316, 1302+356, 1627+289) were not
detected in the 8.4-GHz VLBA observations, and 1425+287 has not been detected in
any VLBA observations. 
Flux densities of the principal components of the sources were measured using the
AIPS task JMFIT and are listed in Table~\ref{table3}.

In addition to the observations described above, unpublished 8.4-GHz VLA
observations of five sources -- 1056+316, 1126+293, 1425+287,
1627+289, 1302+356 -- made in A-conf. by Glen Langston (first four objects)
and \citet{pbww92} have been included 
(Figs.~\ref{1056+316_maps},~\ref{1126+293_maps},~\ref{1425+287_maps},
~\ref{1627+289_maps}, and \ref{1302+356_maps}, respectively).

It was realised that because of poor $u$-$v$ coverage at the higher
frequencies, some flux density could be missing and the resultant spectral index
maps were not considered to be reliable.
Any calculation of spectral indices from the flux densities quoted
in Table~\ref{table3} should also be treated only as coarse approximations.

For 1045+352, 30-GHz continuum observations using the
Toru\'n 32-m radio telescope and a prototype (two-element receiver) 
of the One-Centimeter Receiver Array
\citep[OCRA-p,][]{lowe05} have also been made. The recorded output from the receiver 
was the difference 
between the signals from two closely-spaced horns effectively separated in azimuth
so that atmospheric variations were mostly cancelled out. The observing technique
was such that the respective two beams were pointed at the source alternately with a
switching cycle of $\sim$50 seconds for a period of $\sim$6 minutes, thus measuring the source
flux density relative to the sky background on either side of the source. The telescope
pointing was determined from azimuth and elevation scans across the point source
Mrk\,421.
The primary flux density calibrator that was used was the planetary nebula NGC\,7027,
which has an effective radio angular size of $\sim$8 arcseconds
\citep{bryce97} and for which
a correction of the flux density scale had to be made. However, as NGC\,7027 was at some
distance from the target source, the point source 1144+402 was used as a secondary flux 
density calibrator. Corrections for the effects of the atmosphere were determined from
system temperature measurements at zenith distances of 0$\degr$ and 60$\degr$.

\section{Comments on individual sources}

\begin{figure*}[t]
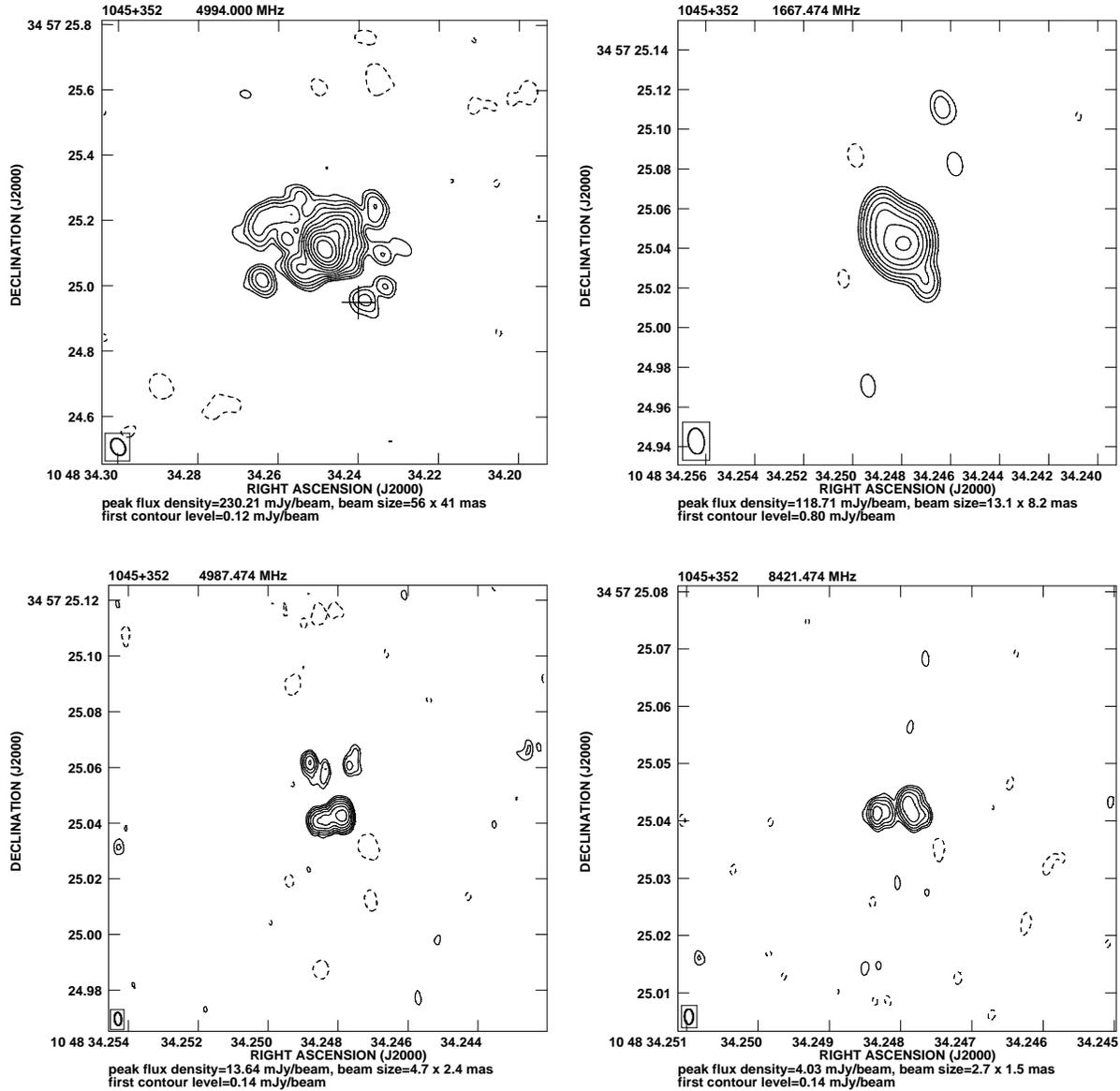

\centering 
\includegraphics[width=8cm, height=8cm]{6364fig1a.ps}
\includegraphics[width=8cm, height=8cm]{6364fig1b.ps}
\includegraphics[width=8cm, height=8cm]{6364fig1c.ps}
\includegraphics[width=8cm, height=8cm]{6364fig1d.ps}
\caption{The MERLIN 5-GHz (upper left) and VLBA 1.7, 5, and 8.4-GHz maps of 
1045+352. Contours increase
by a factor 2, and the first contour level corresponds to $\approx 3\sigma$.
A cross indicates the position of an optical object found using the SDSS/DR5.}
\label{1045+352_maps}
\end{figure*}

\noindent {\bf \object{1045+352}}.
The MERLIN and VLBA maps (Fig.~\ref{1045+352_maps}) show this source to be
extended in both the NE/SW and NW/SE directions. The central compact feature
visible in all the maps is probably a radio core with a steep spectrum. 
The VLBA image at
1.7\,GHz shows two symmetric protrusions -- possibly jets -- straddling
the core in a NE/SW direction, the SW emission being weaker than
in the NE. This structure is aligned with the NE/SW emission visible
in the 5-GHz MERLIN image, but the more extended diffuse emission has been
resolved out in the VLBA images.
The 5-GHz VLBA image shows a core and a one-sided jet pointing to the East.
Some compact features in a NE direction are also visible.
The radio structure in the 8.4-GHz VLBA image is similar to that at
5\,GHz: an extended radio core and a jet pointing in an easterly direction.

The observed radio morphology of 1045+352 could indicate a restart
of activity with the NE/SW radio emission being the first phase of
activity, now fading away, and
the extension in the NW/SE direction being a signature of the current
active phase. However, the above is only one of a number of possible
interpretations of the structure of 1045+352 -- see further discussion in
Sect. 4. 

According to Sloan Digital Sky Survey/Data Release 5 (SDSS/DR5), 1045+352 is a galaxy at
RA=\,$10^{\rm h}48^{\rm m}34\fs242$, Dec=\,$+34\degr 57\arcmin 24\farcs95$,
which is marked with a cross in the MERLIN map but the spectral observations 
carried out by \citet{willott02} have shown 1045+352 to be a quasar with a
redshift of $z=1.604$. It has been also classified as a HiBAL object based
upon the observed very broad C\,IV absorption, and it is a very luminous submillimetre 
object with detections at both 850\,$\mu$m and 450\,$\mu$m \citep{willott02}.

The total flux of 1045+352 at 30\,GHz measured by us using OCRA-p is
$S_{30\mathrm{GHz}}$=69 mJy$\pm7$\,mJy,
which gives a steep spectral index $\alpha=-1.01$  between 4.85\,GHz and
30\,GHz.

\begin{figure*}
\centering
\includegraphics[width=6cm, height=6cm]{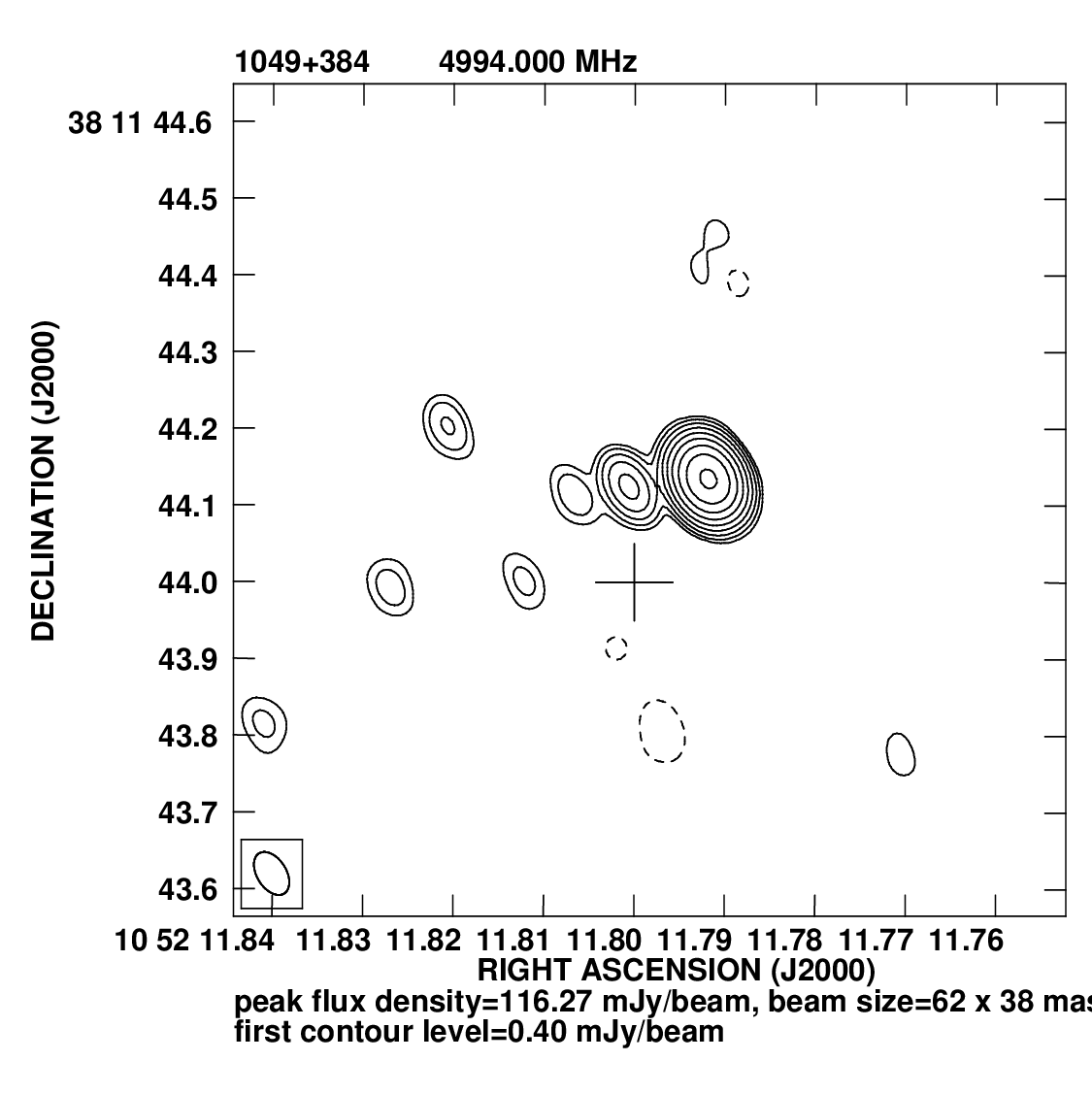}
\includegraphics[width=8cm, height=6cm]{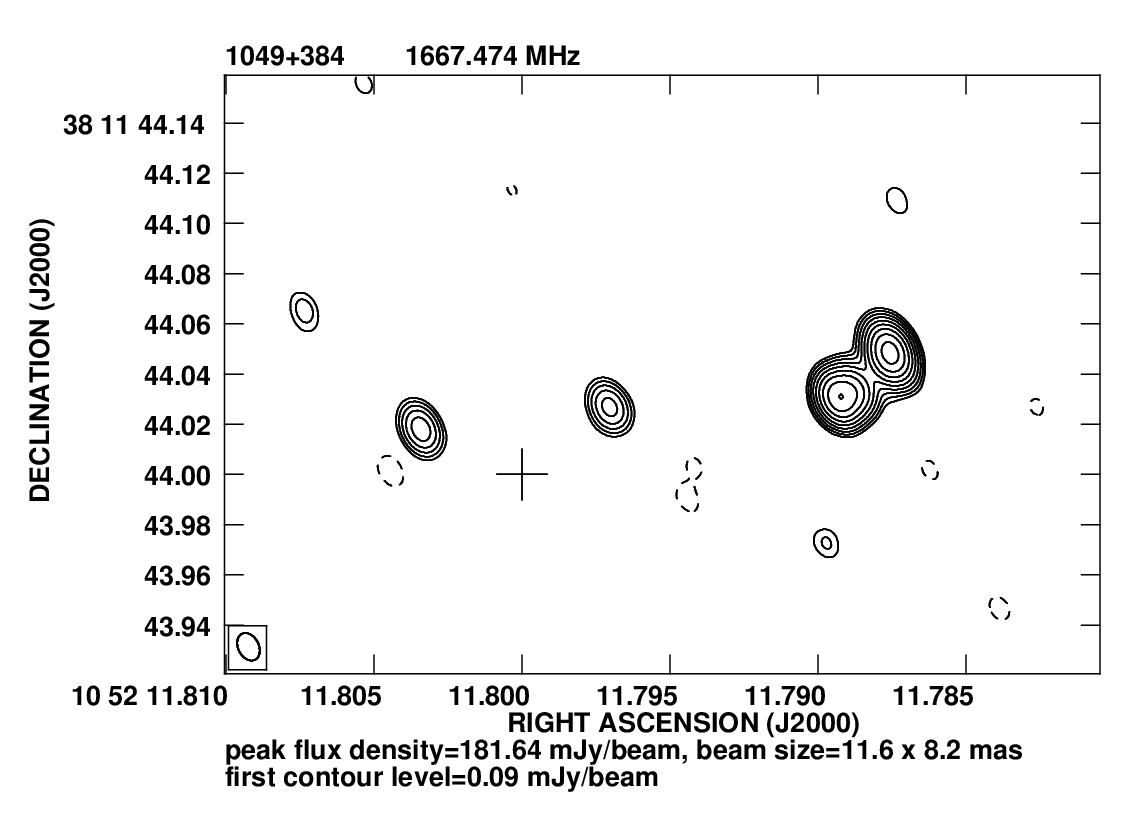}
\includegraphics[width=11cm, height=5.5cm]{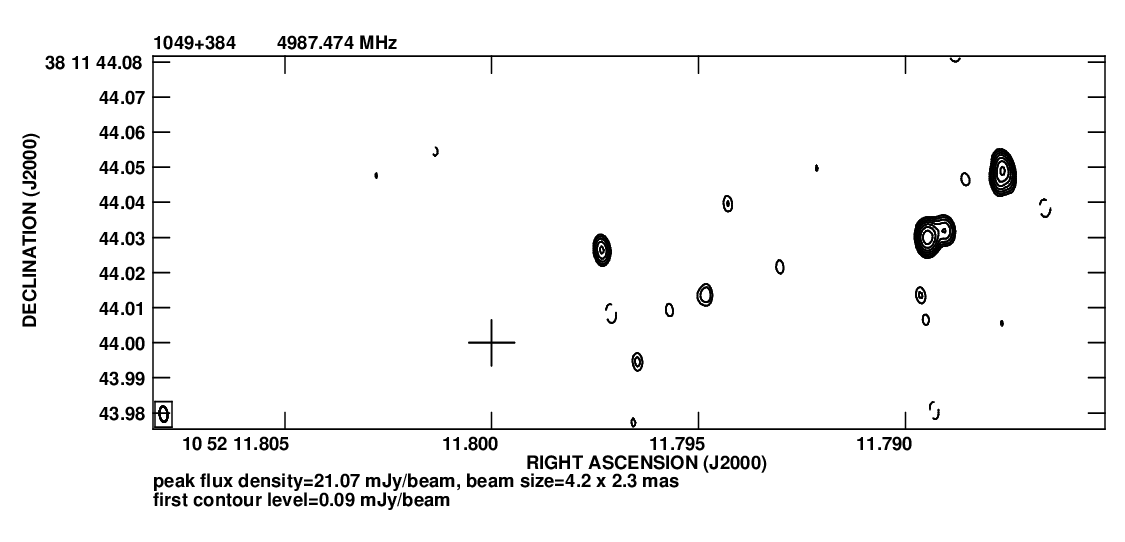}
\includegraphics[width=13cm, height=7.5cm]{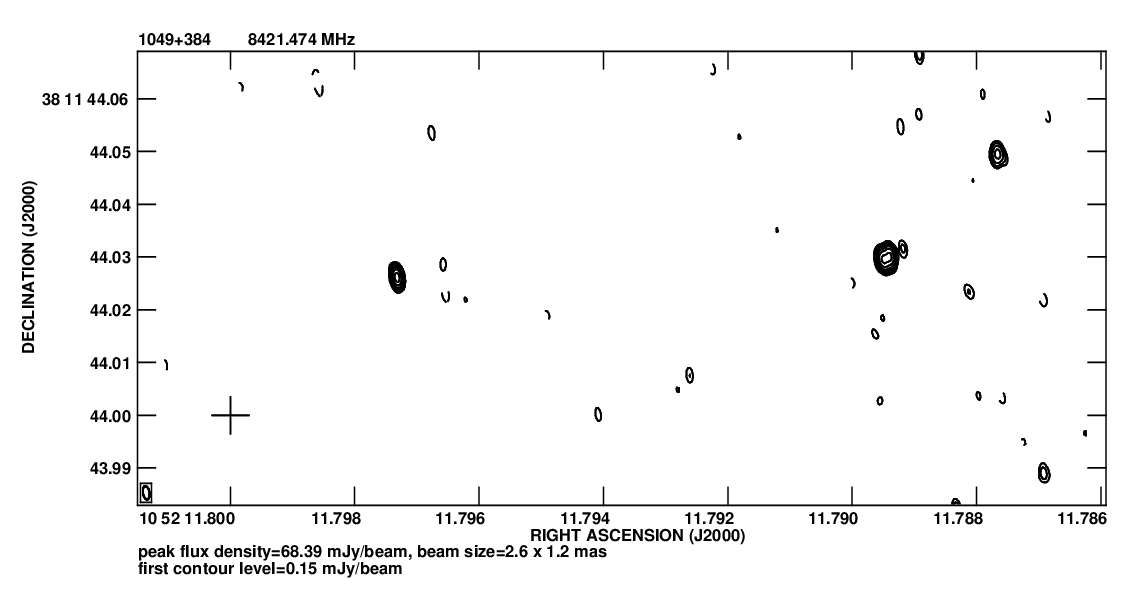}
\caption{The MERLIN 5-GHz (upper left) map and VLBA 1.7, 5, and 8.4-GHz maps of 
1049+384. Contours increase by a factor 2, and the first contour level corresponds
to $\approx 3\sigma$.
Crosses indicate the position of an optical object found using the SDSS/DR5
.}
\label{1049+384_maps}
\end{figure*}

\noindent {\bf \object{1049+384}}.
The 5-GHz MERLIN image (Fig.~\ref{1049+384_maps}) shows it as a triple core-jet structure with the
brightest component resolved into a double structure extended in a NW/SE
direction in the high resolution VLBA observations.
The 1.7-GHz VLBA image shows four radio components \citep[in agreement with
][]{dal02}, whereas the 5-GHz and 8.4-GHz VLBA maps show only three
components. However, the 5-GHz VLBA image published by \citet{or04} shows
all four components, and they suggest that the two western components
and the two eastern ones are two independent radio sources. As
pointed by \citet{or04}, it is difficult to classify the object,
although
the idea that 1049+384 consists of two separate compact, double sources is not
very plausible because of the very small separation, $\sim0.09\arcsec$ (0.4
kpc), between these two potential objects.
Although the spectral index calculations are very uncertain, 
it is suggested that one of the eastern components at RA=\,$10^{\rm h}52^{\rm
m}11\fs797$, Dec=\,$+38\degr 11\arcmin 44\farcs027$ is a radio core
\citep[in agreement with][]{or04} from which jets emerge alternately in opposite
directions.

1049+384 is a galaxy with a redshift $z=1.018$ \citep{rw1994}, but
according to \citet{alli88} the optical spectrum of 1049+384 shows intermediate 
properties between a galaxy and a quasar. The optical
object was included in SDSS/DR5 (RA=\,$10^{\rm h}52^{\rm
m}11\fs802$, Dec=\,$+38\degr 11\arcmin 44\farcs00$) and is marked in all maps with a
cross.

\begin{figure*}
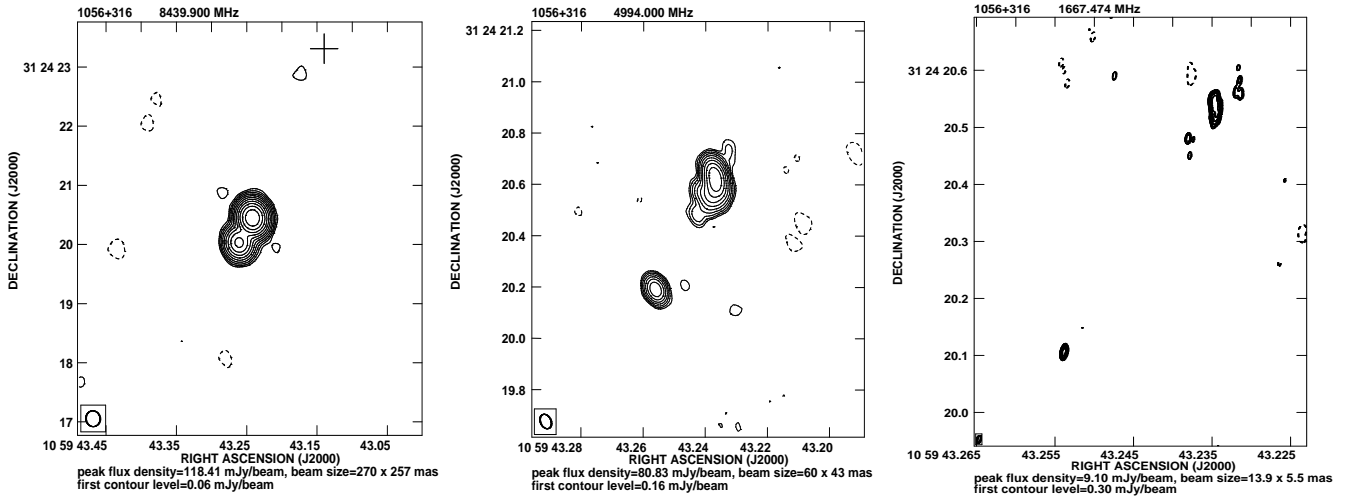

\centering
\includegraphics[width=5.8cm, height=7cm]{6364fig3a.ps}
\includegraphics[width=5.8cm, height=7cm]{6364fig3b.ps}
\includegraphics[width=5.8cm, height=7cm]{6364fig3c.ps}
\caption{The VLA 8.4-GHz map, MERLIN 5-GHz map, and VLBA 1.7-GHz map of
1056+316.
Contours increase by a factor 2, and the first contour level corresponds to
$\approx 3\sigma$. A cross on the VLA map indicates the position of an optical
object found using the SDSS/DR5. }
\label{1056+316_maps}
\end{figure*}

\begin{figure*}
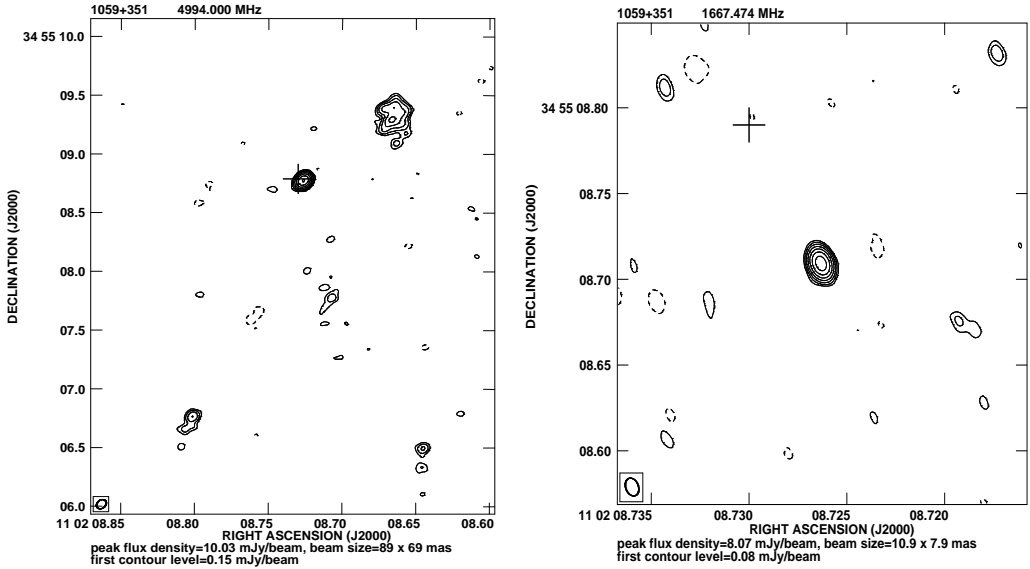

\centering
\includegraphics[width=6.8cm, height=8cm]{6364fig4a.ps}
\includegraphics[width=7cm, height=8cm]{6364fig4b.ps}
\caption{The MERLIN 5-GHz map and VLBA 1.7-GHz map of 1059+351.
Contours increase by a factor 2, and the first contour level corresponds
to $\approx 3\sigma$. Crosses indicate the position of an optical object
found
using the SDSS/DR5.}
\label{1059+351_maps}
\end{figure*}

\noindent {\bf \object{1056+316}}.
The 8.4-GHz VLA image (Fig.~\ref{1056+316_maps}) shows this source to have 
a double structure that,
in the 5-GHz MERLIN image, has been resolved into a radio core and probably a hotspot 
in a NW radio lobe. Both components are visible in the 1.7-GHz VLBA image,
but neither has been detected in the higher frequency VLBA images.
The two weak features on either side of the NW component  in the 1.7-GHz
VLBA image may be the remains of extended emission that has been resolved out.

The optical counterpart of 1056+316 was included in SDSS/DR5
(RA=\,$10^{\rm h}59^{\rm m}43\fs145$, Dec=\,$+31\degr 24\arcmin 23\farcs31$), 
together with a photometric redshift (Table~\ref{table1}).
Its position is marked with a cross in 8.4-GHz VLA map.

\begin{figure*}
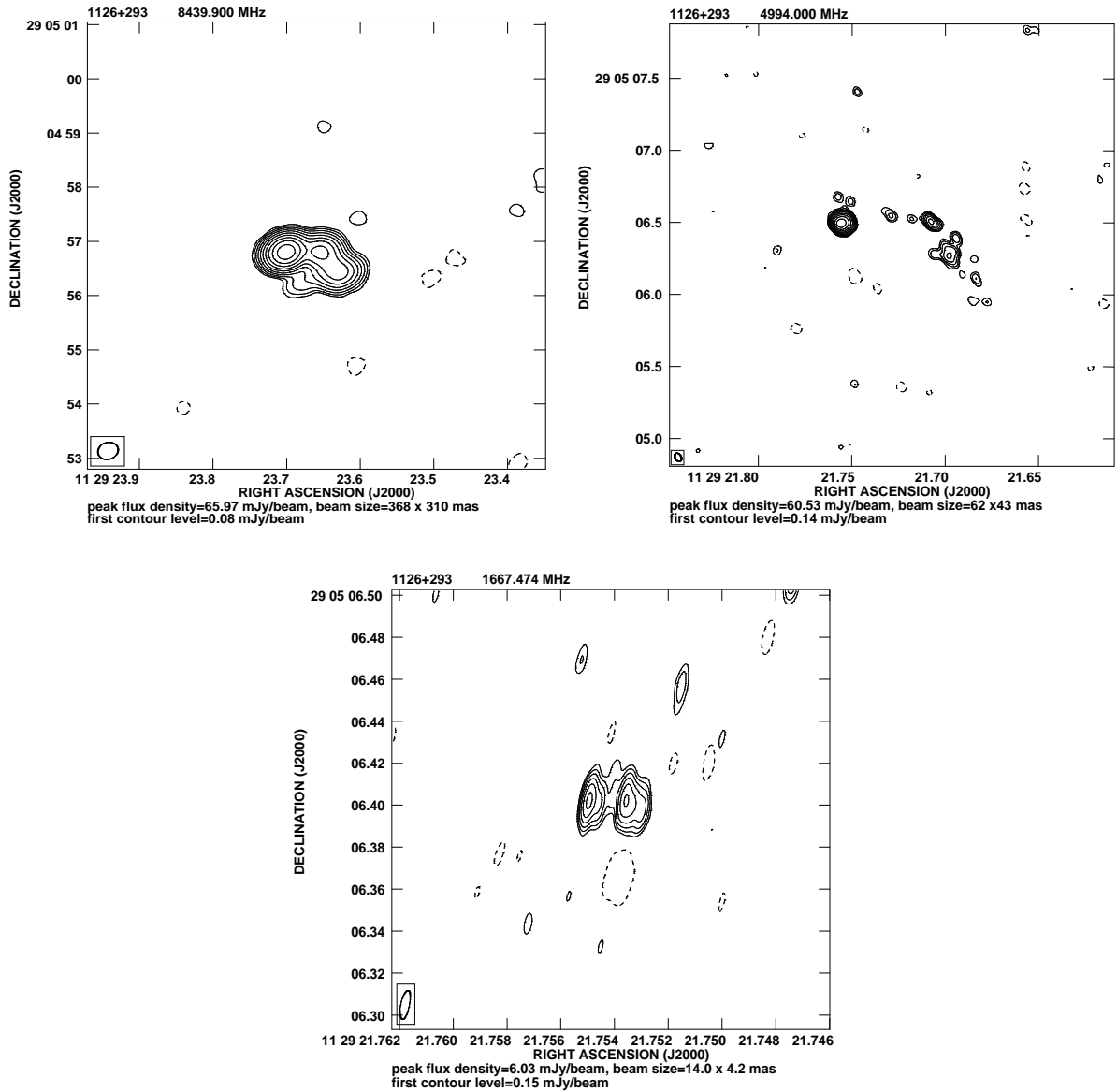

\centering
\includegraphics[width=8cm, height=8cm]{6364fig5a.ps}
\includegraphics[width=8cm, height=8cm]{6364fig5b.ps}
\includegraphics[width=8cm, height=8cm]{6364fig5c.ps}
\caption{The VLA 8.4-GHz map, MERLIN 5-GHz map and VLBA 1.7-GHz map of
1126+293.
Contours increase by a factor 2, and the first contour level corresponds
to $\approx 3\sigma$.}
\label{1126+293_maps}
\end{figure*}

\noindent {\bf \object{1059+351}}.
The 5-GHz MERLIN map (Fig.~\ref{1059+351_maps}) shows a 
bright component that is probably a radio core, on almost opposite sides of
which is emission from compact features (hotspots) within the two radio
lobes. This structure
agrees with the 1.4-GHz VLA  observations presented by
\citet{greg88} and  \citet{mach83}. Their images clearly show an S-shaped
morphology of 1059+351 with two very diffuse components, the brighter
one resolved into a double structure in 5-GHz VLA
observations
\citep{mach98}. One of these two components is the NW hotspot visible in the
5-GHz MERLIN map, and the second is probably a radio core visible in both the 5-GHz
MERLIN and 1.7-GHz VLBA images.
 
The optical counterpart of 1059+351 was included in SDSS/DR5
(RA=\,$11^{\rm h}02^{\rm m}08\fs727$, Dec=\,$+34\degr 55\arcmin
08\farcs79$), together with a photometric redshift (Table~\ref{table1}). 
The position of the optical
object is marked with a cross in all maps and is well correlated
with the position of the radio core. \citet{mach98} also measured a photometric
redshift for 1059+351, which is $z=0.37$ and which differs from that in SDSS/DR5.

\noindent {\bf \object{1126+293}}. 
The VLA 8.4-GHz and MERLIN 5-GHz maps (Fig.~\ref{1126+293_maps}) show three radio components,
the brighter one probably being the core that was resolved into a
core-jet structure in the 1.7-GHz VLBA image.
The source was not detected in the 5 and 8.4-GHz VLBA observations.

\begin{figure*}
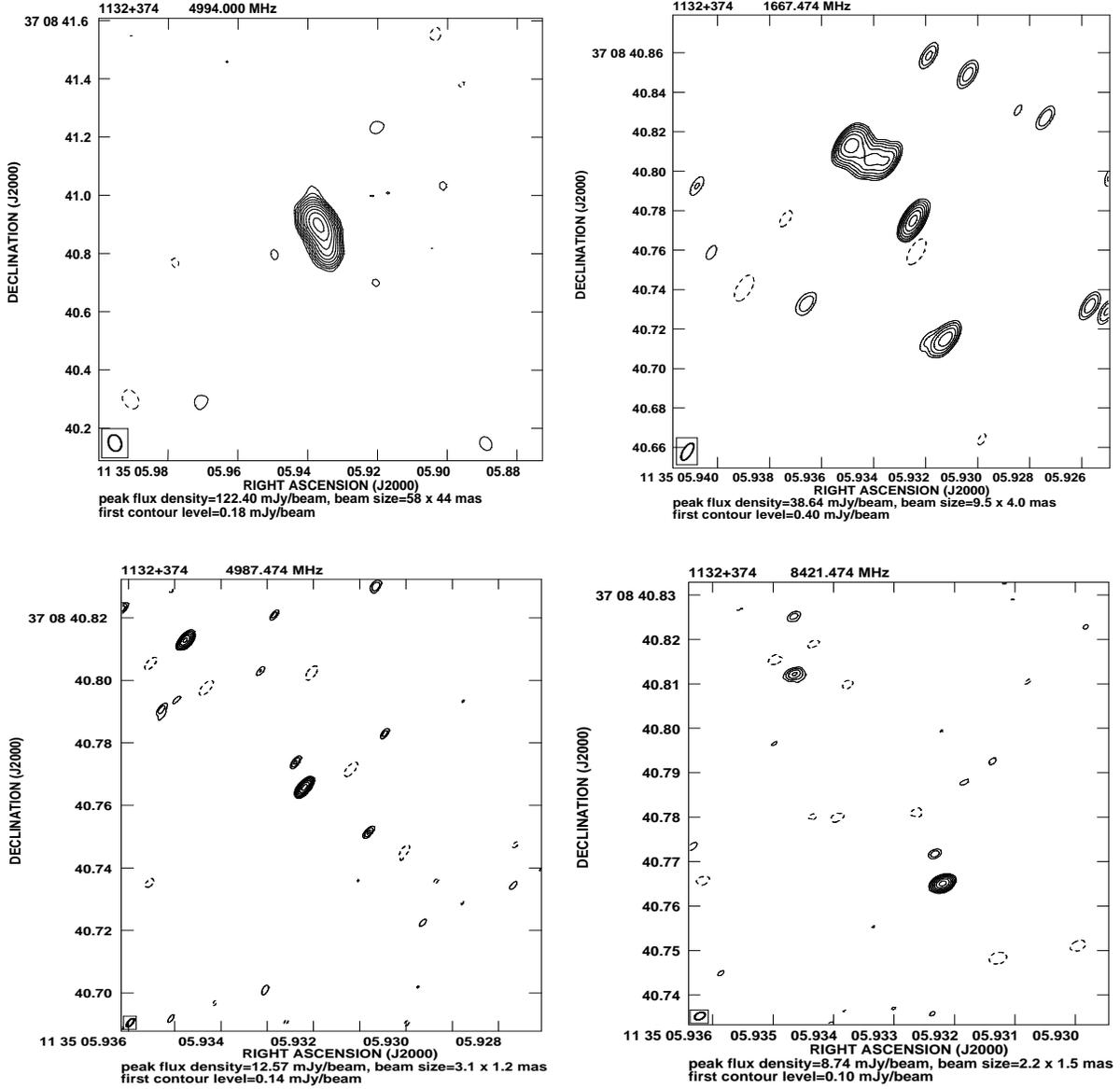

\centering
\includegraphics[width=8cm, height=8cm]{6364fig6a.ps}
\includegraphics[width=8cm, height=8cm]{6364fig6b.ps}
\includegraphics[width=8cm, height=8cm]{6364fig6c.ps}
\includegraphics[width=8cm, height=8cm]{6364fig6d.ps}
\caption{The MERLIN 5-GHz (upper left) map and VLBA 1.7, 5, and 8.4-GHz maps of 1132+374.
Contours increase by a factor 2, and the first contour level corresponds
to $\approx 3\sigma$.}
\label{1132+374_maps}
\end{figure*}

\begin{figure*}
\centering
\includegraphics[width=8cm, height=8cm]{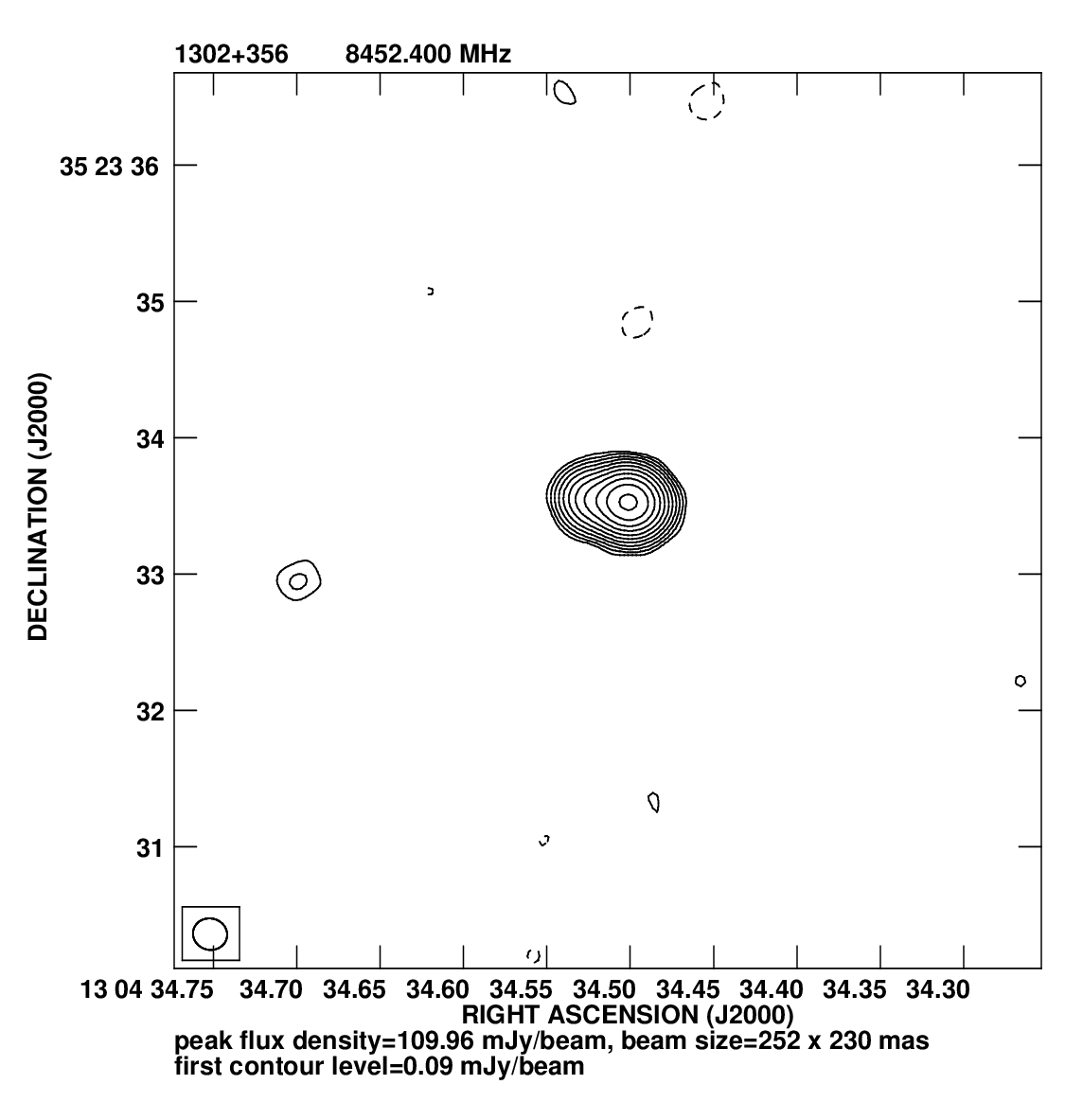}
\includegraphics[width=8cm, height=8cm]{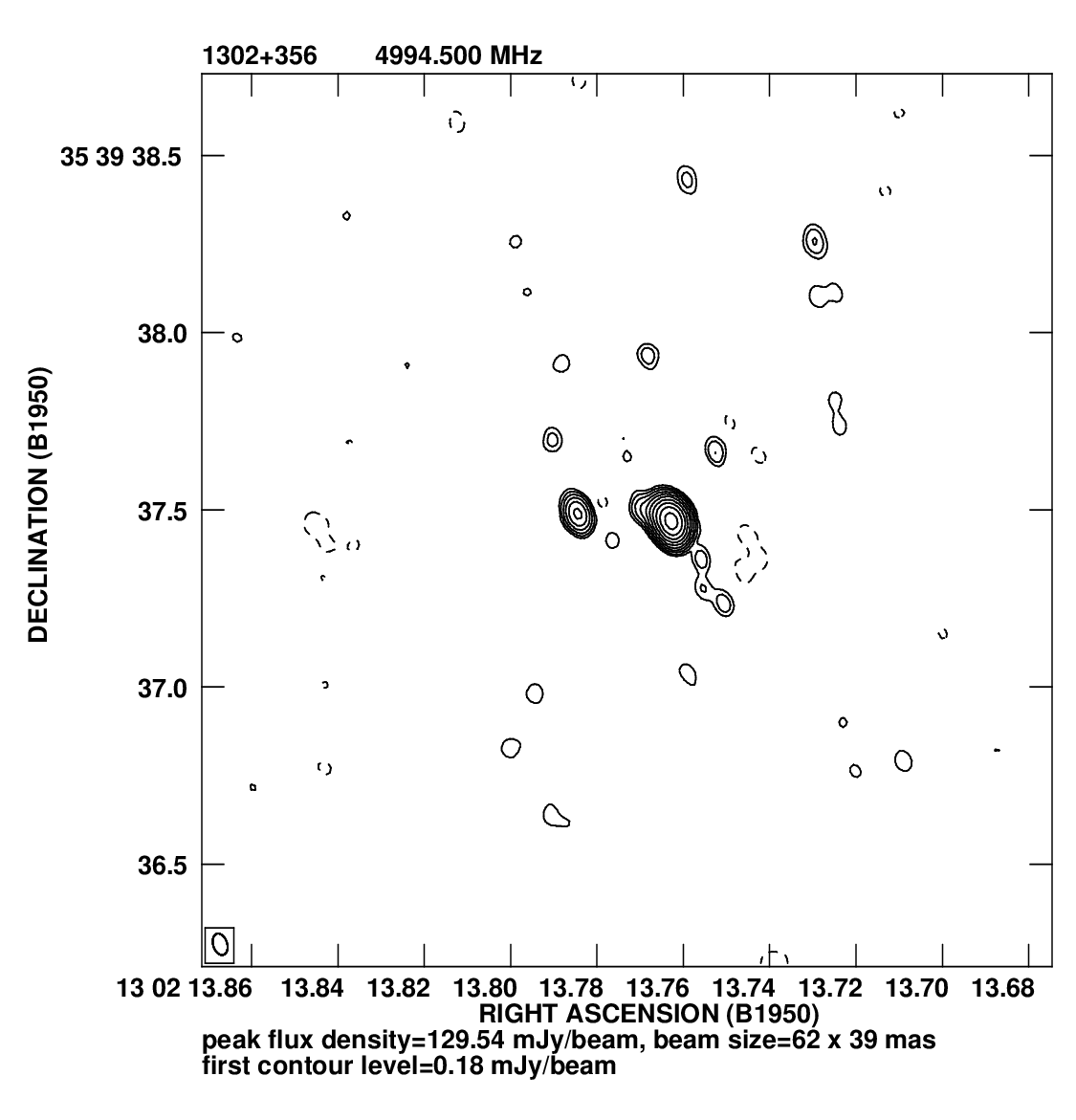}
\includegraphics[width=8cm, height=8cm]{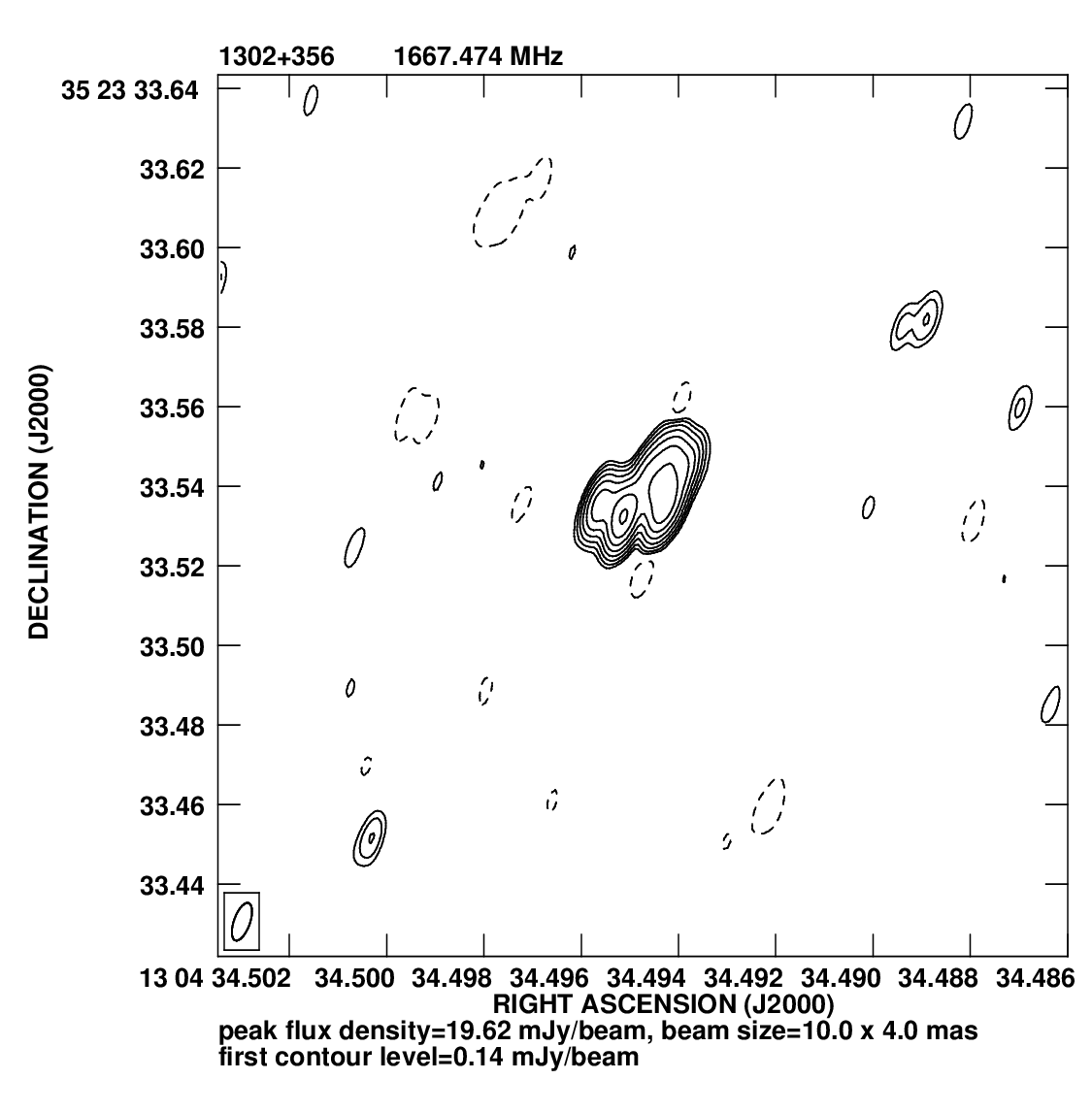}
\includegraphics[width=8cm, height=8cm]{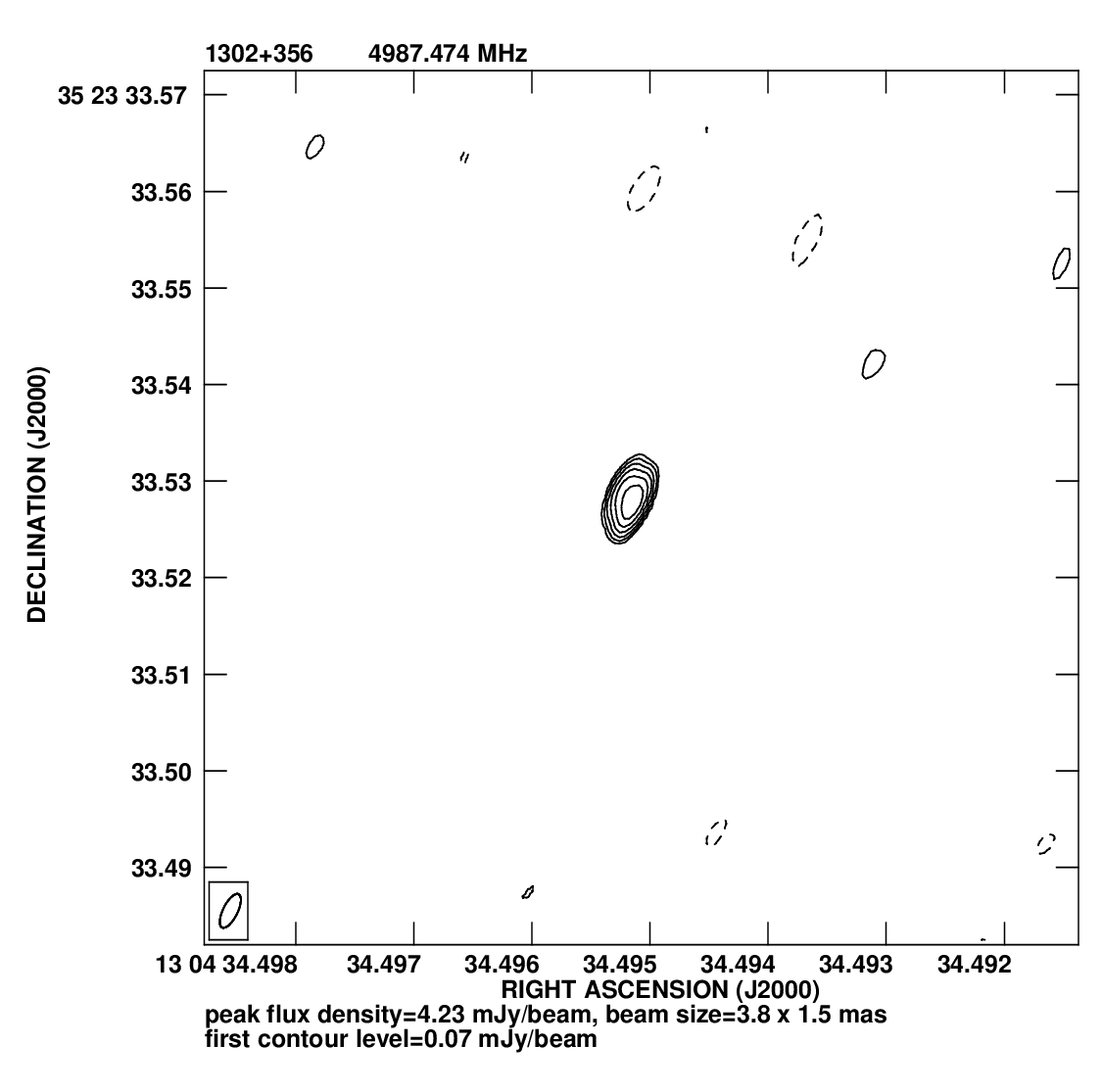}
\caption{The VLA 8.4-GHz map (upper left), MERLIN 5-GHz map (upper right) 
and VLBA 1.7 and 5-GHz maps of 1302+356.
Contours increase by a factor 2, and the first contour level corresponds
to $\approx 3\sigma$.}
\label{1302+356_maps}
\end{figure*}

\noindent {\bf \object{1132+374}}.
The 5-GHz MERLIN image shows (Fig.~\ref{1132+374_maps}) a core-jet structure
that was resolved into a triple CSO object in the 1.7-GHz VLBA image.
The 5 and 8.4-GHz VLBA images show only two components: a hotspot in the
NE lobe and a radio core.
This source is identified with a very high redshift ($z=2.88$) galaxy \citep{er1996}.

\noindent {\bf \object{1302+356}}.
This source was observed with the VLA at 8.4\,GHz as a part of the JVAS
survey
\citep{pbww92}. The resulting map shows a slightly extended EW
object (Fig.~\ref{1302+356_maps}). The 5-GHz MERLIN image shows
this to be a double source, and the weak ($\sim$10 mJy) eastern component
could be part of a jet. The bright component was resolved into a diffuse
structure in the 1.7-GHz VLBA image. The 5-GHz VLBA image shows only a single
component at the position of the maximum emission in the 1.7-GHz VLBA image,
which is probably a radio core (Fig.~\ref{1302+356_maps}). There is no trace
of this source in the 8.4-GHz VLBA image.

\begin{figure*}
\centering
\includegraphics[width=8cm, height=8cm]{6364fig8a.ps}
\includegraphics[width=8cm, height=8cm]{6364fig8b.ps}
\includegraphics[width=8cm, height=8cm]{6364fig8c.ps}
\includegraphics[width=8cm, height=8cm]{6364fig8d.ps}
\caption{The MERLIN 5-GHz map (upper left) and VLBA 1.7, 5, and 8.4-GHz maps of 1407+369.
Contours increase by a factor 2, and the first contour level corresponds
to $\approx 3\sigma$. Crosses indicate the position of an optical object
found
using the SDSS/DR5.}
\label{1407+369_maps}
\end{figure*}

\begin{figure*}
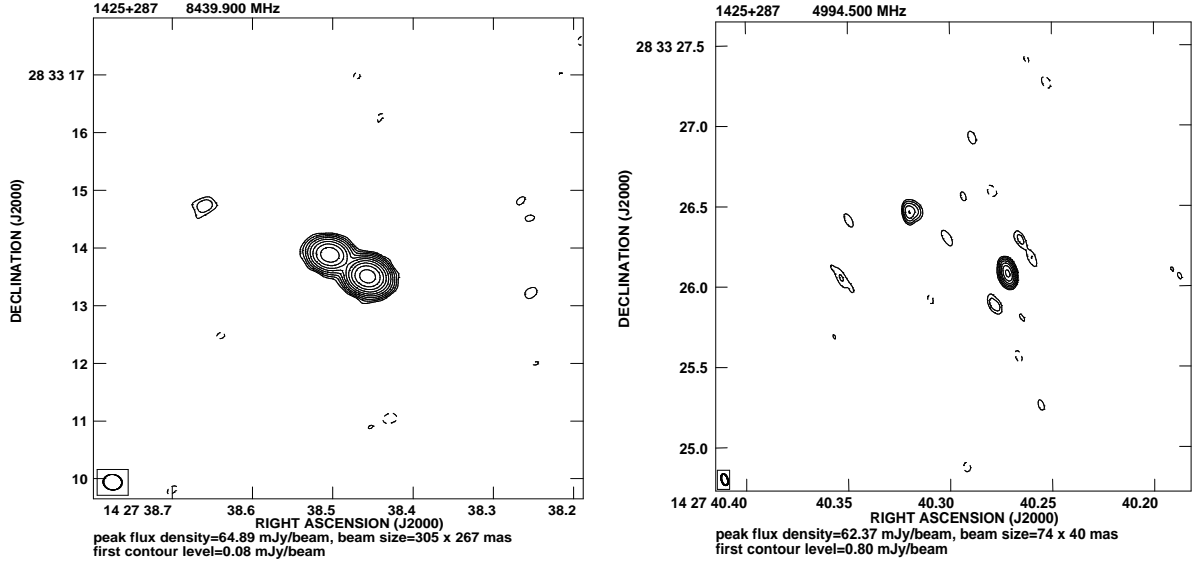

\centering
\includegraphics[width=8cm, height=8cm]{6364fig9a.ps}
\includegraphics[width=8cm, height=8cm]{6364fig9b.ps}
\caption{The VLA 8.4-GHz map and MERLIN 5-GHz map of 1425+287.
Contours increase by a factor 2, and the first contour level corresponds
to $\approx 3\sigma$.}
\label{1425+287_maps}
\end{figure*}

\begin{figure*}
\centering
\includegraphics[width=8cm, height=8cm]{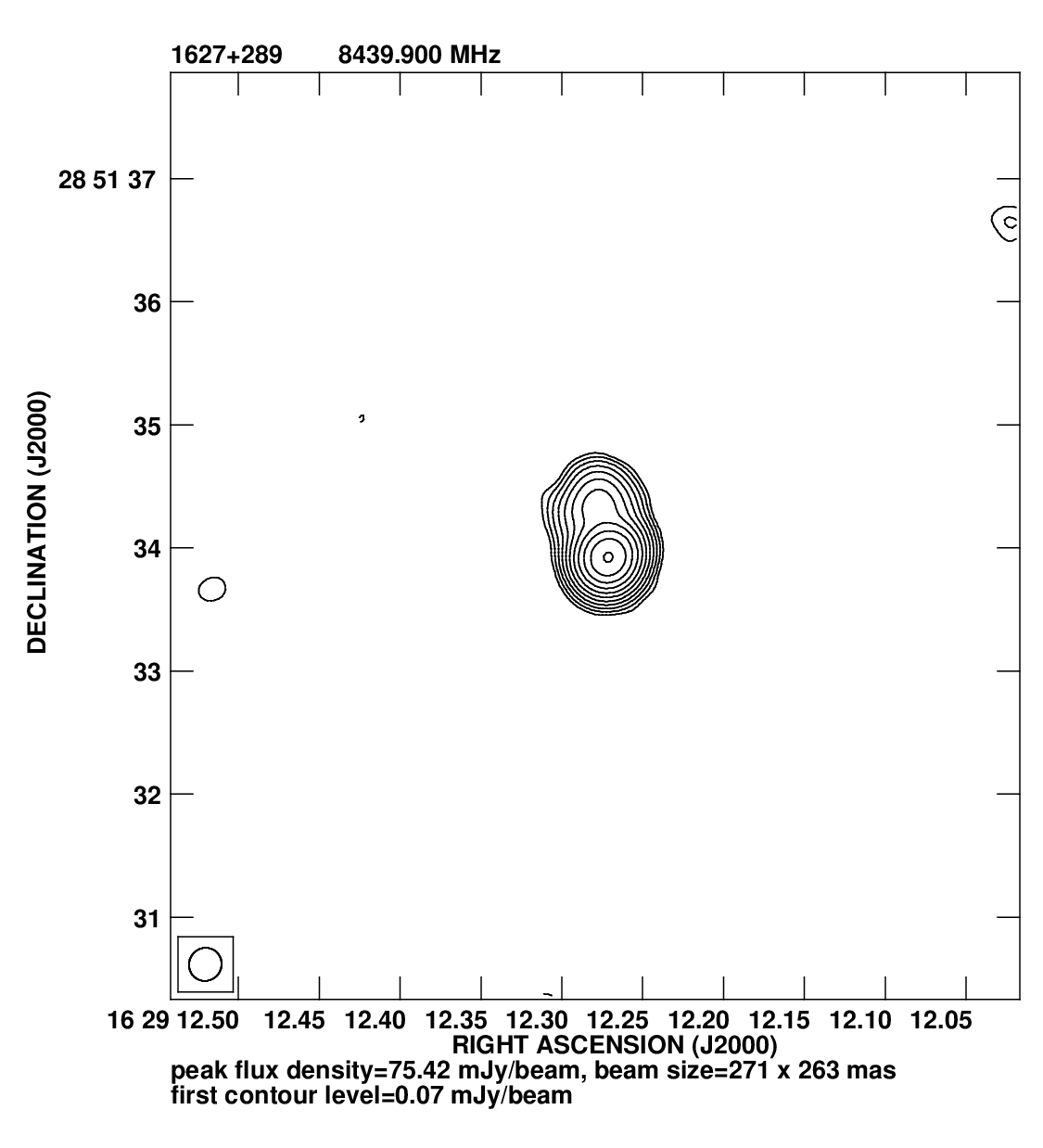}
\includegraphics[width=8cm, height=8cm]{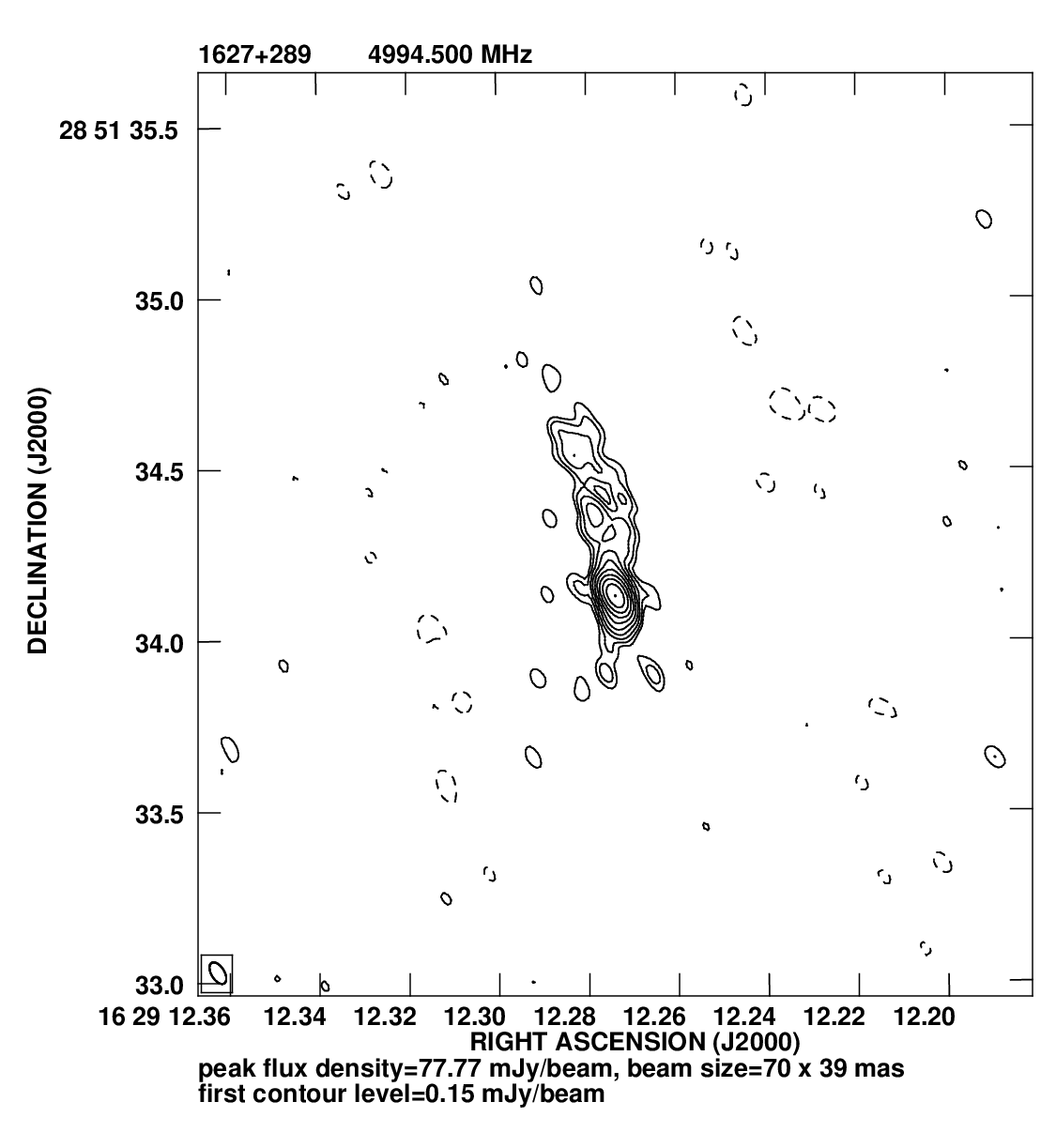}
\includegraphics[width=8cm, height=8cm]{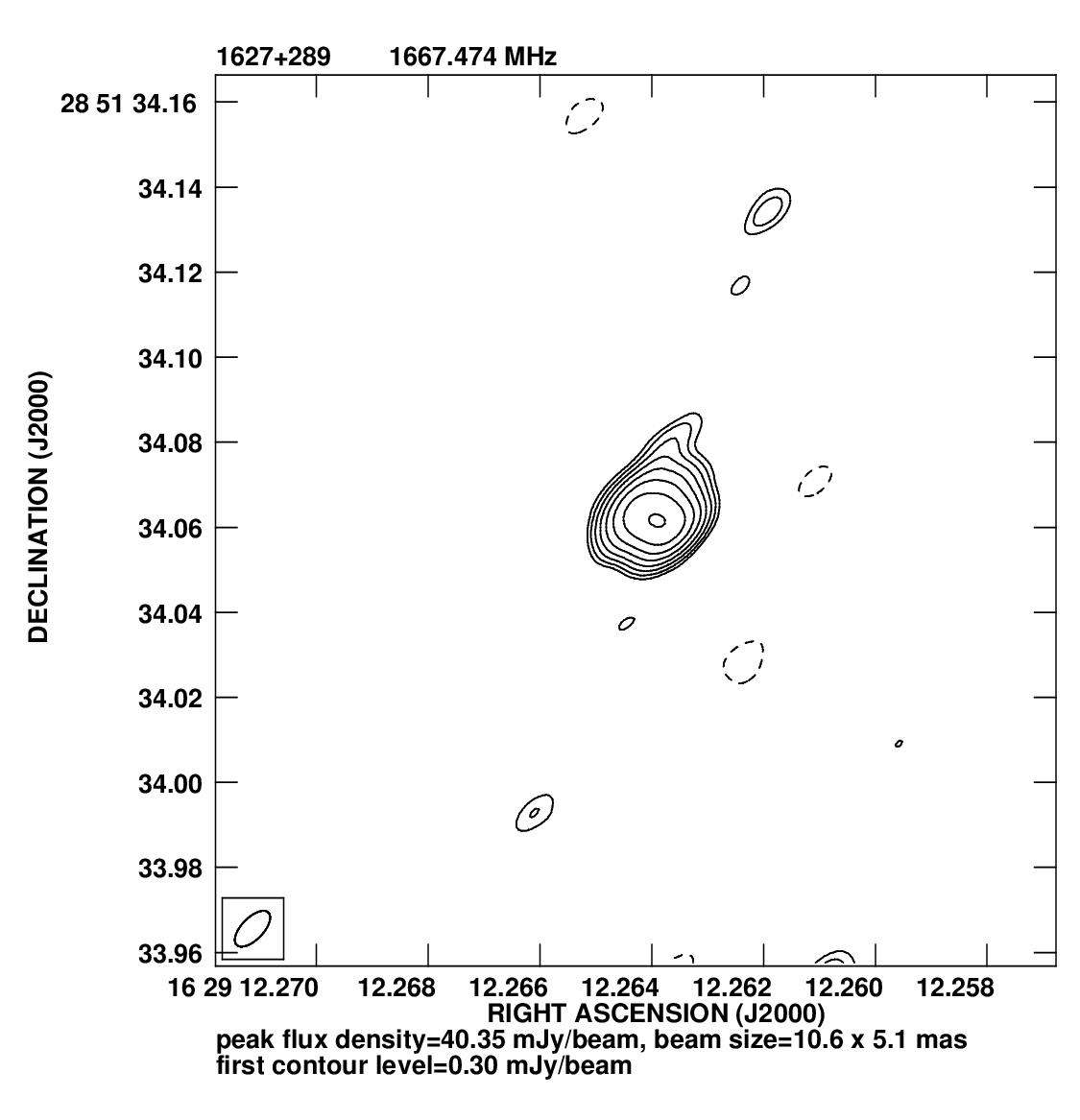}
\includegraphics[width=8cm, height=8cm]{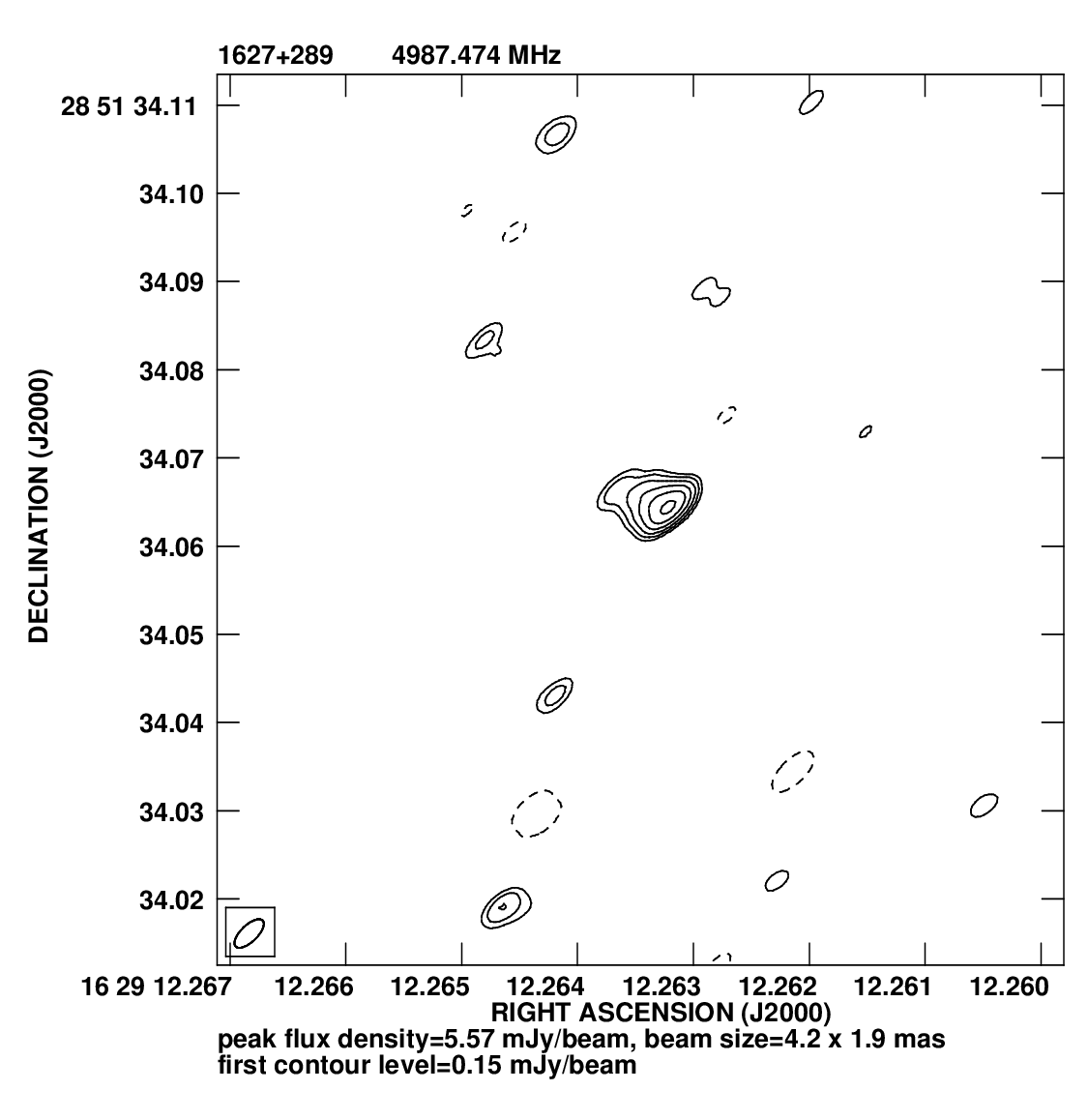}
\caption{The VLA 8.4-GHz map (upper left), MERLIN 5-GHz map (upper right), and VLBA 1.7 and 5-GHz maps
of 1627+289.
Contours increase by a factor 2, and the first contour level corresponds
to $\approx 3\sigma$.}
\label{1627+289_maps}
\end{figure*}

\noindent {\bf \object{1407+369}}.
The 5-GHz MERLIN image
shows a core-jet structure in a NW direction that is
resolved into
a core and jet in all the VLBA maps (Fig.~\ref{1407+369_maps}).
The optical object was included in SDSS/DR5 (RA=\,$14^{\rm h}09^{\rm
m}09\fs509$,
Dec=\,$+36\degr 42\arcmin 08\farcs15$) and is marked with a cross in all maps.
The redshift quoted in Table~\ref{table1} is photometric.

\noindent {\bf \object{1425+287}}.
Both the VLA 8.4-GHz and MERLIN 5-GHz images (Fig.~\ref{1425+287_maps})
show a double structure for this source. The brighter component seems
to be a radio core, although this cannot be confirmed because the source was
not detected in the VLBA observations (Fig.~\ref{1425+287_maps}).

\noindent {\bf \object{1627+289}}.
Both the VLA 8.4-GHz and MERLIN 5-GHz images (Fig.~\ref{1627+289_maps}) show
this source to have a core-jet structure. The 1.7-GHz VLBA image shows only the central
extended feature that was resolved into a core-jet
structure in the 5-GHz VLBA image. The source was
not detected in the 8.4-GHz VLBA image.

\section{Discussion}
\subsection{1045+352 --- a BAL quasar}

1045+352 is a HiBAL quasar with a very reddened spectrum showing a
C\,IV broad absorption system \citep{willott02}. 
Its projected linear size is only 2.1 kpc, which is consistent with 
the observation of \citet{becker00}
that, amongst radio loud quasars, broad absorption lines are more commonly
observed in the smallest radio sources.   

\begin{figure}[t]
\includegraphics[width=6cm, height=8.5cm,
angle=-90]{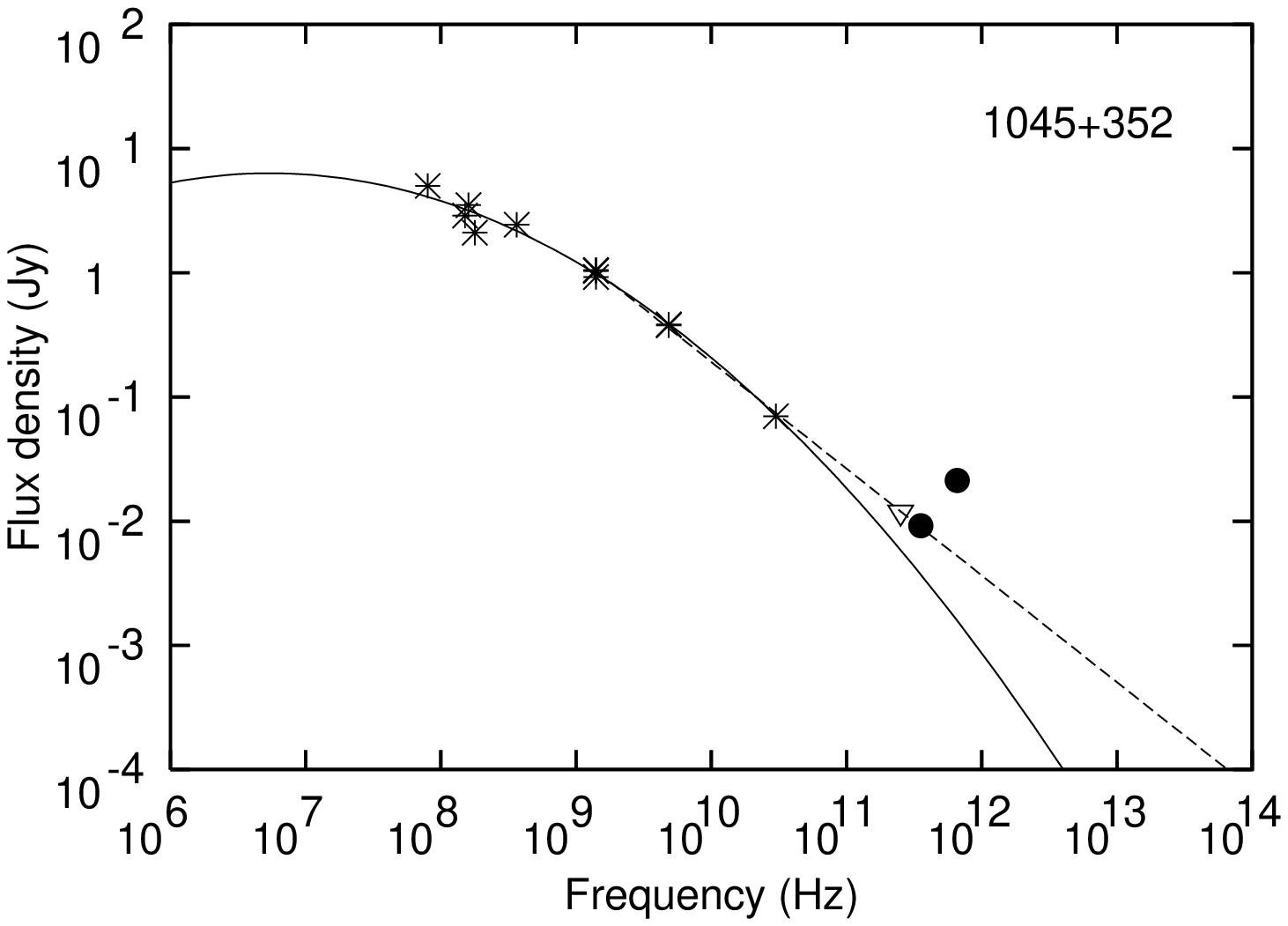}
\caption{Spectral Energy Distribution (SED) of 1045+352 from radio to submillimetre
wavelengths. The errors are smaller than the size of the symbols; 1.25\,mm point 
\citep{haas06} is shown as a triangle, 850\,$\mu$m
and 450\,$\mu$m points \citep{willott02} are shown as filled circles, radio
observations are shown as asterisks. The solid curve is the parabolic fit
$f(x)=ax^{2}+bx+c$ to all radio data ($y_{i}$), with $a=-0.14,
b=1.91, c=-5.68$, and reduced $\chi^{2}=12$. 
The dashed curve is the linear fit $f(x)=ax+b$ to radio data with
$\nu>1 {\rm GHz}$, with $a=-0.86, b=7.91$, and reduced
$\chi^{2}=0.5$.}
\label{plot_maps}
\end{figure}

It is a very luminous submillimetre object, which together with the template
dust spectrum adopted by \citet{willott02}, indicates this source to be a hyperluminous
infrared quasar, with large amounts of dust in its host galaxy.
Although 1045+352 is quite luminous at 151\,MHz \citep[2.88\,Jy,][]{wal96}, which 
suggests the
presence of some extended emission and which, indeed, appears to be present in our 
MERLIN 5-GHz image, the VLBA maps show the radio structure to be dominated by jets and a core.   
The 30-GHz flux density of 1045+352 is also high, as would be expected from the VLBA structure. 
Consequently, there could be synchrotron contamination of the submillimetre flux. 
As shown by \citet{blun99}, either the first-order or second-order
polynomials can accurately predict the shape
of the radio spectrum. Both models have been applied to the radio data
of 1045+352 taken from the literature and from this paper (Fig.~\ref{plot_maps}), and 
show that a non-thermal component could constitute at least $\sim$40$\%$
of the entire 850\,$\mu$m flux (the parabolic fit). The linear fit agrees 
with calculations based upon the 1.25\,mm flux measured by \citet{haas06}, who derived 
a value of 94$\%$ for the non-thermal component part of the detected 850\,$\mu$m flux. 
It has to be noted here that the linear fit should be treated as
an upper limit for the synchrotron emission at submillimetre wavelengths, 
since the spectrum may steepen in the interval between 30\,GHz and the SCUBA wavebands. 
However, the above can indicate values of infrared
emission and dust mass of 1045+352 lower than estimated \citep{willott02}.
This also appears be consistent with the findings of
\citet{willott03}, who have shown that there is no difference between the
submillimetre luminosities of BAL and non-BAL quasars, which suggest that 
a large dust mass is not required for quasars to show BALs.

\begin{table}[h]
\caption[]{1045+352 properties}
\begin{tabular}{@{}l c@{}}
\hline
\hline
~~~~~Parameter & Value~~~~~\\
\hline
~~~~~$u'$     & 22.12~~~~~\\
~~~~~$g'$     & 21.38~~~~~\\
~~~~~$r'$     & 20.81~~~~~\\
~~~~~$i'$     & 20.14~~~~~\\
~~~~~$z'$     & 20.08~~~~~\\
~~~~~$A_{B}$     & 2.0~~~~~\\
~~~~~$M_{B}$     & -22.05 (-24.05)~~~~~\\
~~~~~$A_{V}$     & 1.5~~~~~\\
~~~~~$M_{V}$     & -22.83 (-24.33)~~~~~\\
 ~~~~~$\log(R^{\ast})$(total) & 4.9 (4.1)~~~~~\\
~~~~~$\log(R^{\ast})$(core) & 3.8 (3.0)~~~~~\\
\hline
\end{tabular}

\vspace{0.5cm}
{\small
Notes: Optical photometry from SDSS, corrected for Galactic
extinction. $A_{V}$ taken from \citet{willott02}.
Quantities in parentheses are
corrected for intrinsic extinction. 
}
\label{table2}
\end{table}

The radio luminosity at 1.4\,GHz is high (Table~\ref{table1}), making this
source one of the most radio-luminous BAL quasars, with a value
similar to that of the first known radio-loud BAL\,QSO with an FR\,II
structure, FIRST\,J101614.3+520916 \citep{gregg00}.
Following \citet{stocke92}, a radio-loudness parameter,
$R^{\ast}$, defined as the {\it K}-corrected ratio of the 5-GHz radio flux
to 2500$\AA$ optical flux (Table~\ref{table2}) was calculated. 
For this, a global radio spectral index, $\alpha_{radio}=-0.8$ and an optical
spectral index, $\alpha_{opt}=-1.0$, were assumed, and the SDSS 
$g'$ magnitude defined by \citet{fuku96} was converted to the
Johnson-Morgan-Cousins {\it B} magnitude using the formula given by
\citet{smith02}. Corrections were also made for intrinsic extinction 
(local to the quasar) calculated by \citet{willott02}, who assumed a Milky-Way 
extinction curve. Even after correction, $\log(R^{\ast})>1$, which means 
that 1045+352 is still radio-loud object. 
The angle between the jet axis and the line of sight can be estimated
using the core radio-to-optical luminosity ratio defined by
\citet{wills95} as $\log(R_{V})=\log(L_{core})+0.4M_{V}-13.69$, where 
$L_{core}$ is a radio luminosity of the core at 5-GHz rest frequency
(the core flux density at 5\,GHz were taken from the VLBA image; see also
Table~\ref{table3}), and $M_{V}$ is the {\it K}-corrected absolute magnitude 
calculated using 
transformation equation $V=g'-0.55(g'-r')-0.03$ \citep{smith02}. 
From this, a value of $\sim$3.2 has been obtained for 1045+352, implying an 
angle in the range $\theta\sim10\degr-30\degr$ for the jet 
in the observed asymmetric MERLIN 5-GHz radio morphology, and can explain the 
high value of the radio-loudness parameter. 
An assumption of $\theta=20\degr$ yields the deprojected linear 
size of the source of $\sim6$~kpc. 
As shown by \citet{white06}, BAL\,QSOs are 
systematically brighter than non-BAL objects, which indicates we are looking
closer to the jet axis in quasars with BALs.
Based upon the small inclination angles of their
BAL quasars, \citet{zhou06} suggest that BAL features can be caused
by polar disk winds. Also, \citet{saikia01}
and \citet{jeyakumar05} found that the radio properties of CSS sources are
consistent with the unified scheme in which the axes of the quasars are observed 
close to the line of sight. On the other hand, it has been shown
\citep{saikia01,jeyakumar05} that many CSS objects interact with an asymmetric
medium in the central regions of their host galaxies, and this can cause the observed 
asymmetries. It is then likely that, also in the case of the CSS quasar
1045+352, the environmental asymmetries might play an important role. 
The jet power can be
estimated from the relationship between the radio luminosity and the jet power 
given by \citet[][Eq.(12)]{willott99}. 
However, because some of the flux density of the 1045+352 can be
beamed, the calculations have to be treated as an approximation. 
Assuming the 151-MHz flux density, which accounts for the extended emission
and the radio emission from the jets, the jet kinetic power is 
$Q_{jet}\sim10^{44} {\rm erg~sec^{-1}}$.

\begin{table*}[t]
\caption[]{Flux densities of sources principal components from the VLBA
observations}
\begin{center}
\begin{tabular}{@{}c c c c c c c c l@{}}
\hline
\hline
~~~Source & RA & DEC &
\multicolumn{1}{c}{${\rm S_{1.7\,GHz}}$}&
\multicolumn{1}{c}{${\rm S_{5\,GHz}}$}&
\multicolumn{1}{c}{${\rm S_{8.4GHz}}$}&
\multicolumn{1}{c}{$\theta_{1}$}&
\multicolumn{1}{c}{$\theta_{2}$}&
\multicolumn{1}{c}{PA}~~~\\
~~~Name   & h~m~s & $\degr$~$\arcmin$~$\arcsec$ &
\multicolumn{1}{c}{mJy}&
\multicolumn{1}{c}{mJy}&
\multicolumn{1}{c}{mJy}&
\multicolumn{1}{c}{mas}&
\multicolumn{1}{c}{mas}&
\multicolumn{1}{c}{$\degr$}~~~\\
~~~(1)& (2) & (3) &
\multicolumn{1}{c}{(4)}&
\multicolumn{1}{c}{(5)}&
\multicolumn{1}{c}{(6)}&
\multicolumn{1}{c}{(7)}&
\multicolumn{1}{c}{(8)}&
\multicolumn{1}{c}{(9)~~~}\\
\hline
1045+352&10 48 34.248&34 57 25.044&303.2&$-$ &$-$&15.0&11.0&60\\
        &10 48 34.249&34 57 25.061&$-$&3.5 &$-$&2.0&1.0&76\\
        &10 48 34.248&34 57 25.041&$-$&21.8&7.1&7.0&1.0&101\\
        &10 48 34.248&34 57 25.043&$-$&32.7&12.3&4.0&3.0&95\\
\hline
1049+384&10 52 11.803&38 11 44.018&13.6 &$-$ &$-$ &3.0&1.0&14\\
        &10 52 11.797&38 11 44.027&11.4 &3.9 &6.9 &2.0&2.0&121\\
        &10 52 11.789&38 11 44.031&182.1&33.6&12.9&8.0&1.0&119\\
        &10 52 11.787&38 11 44.048&218.5&23.9&2.3 &5.0&3.0&177\\
\hline
1056+316&10 59 43.254&31 24 20.106&8.8 &$-$ &$-$ &0.9&0.1 &7\\
        &10 59 43.235&31 24 20.538&43.6&$-$ &$-$ &33.0 &8.0&6\\
\hline
1059+351&11 02 08.726&34 55 08.709&8.1&$-$ &$-$ &0.7&0.3 &124\\
\hline
1126+293&11 29 21.755&29 05 06.402&7.3&$-$ &$-$ &3.0&1.0&84\\
        &11 29 21.753&29 05 06.401&10.4&$-$ &$-$ &13.0&4.0&53\\
\hline
1132+374&11 35 05.934&37 08 40.810&124.1&6.6  &1.6&18.0&2.0&57\\
        &11 35 05.932&37 08 40.775&36.3 &13.8 &9.4&2.0&0.4&8\\
        &11 35 05.931&37 08 40.715&14.5 &$-$  &$-$&5.0&0.8&105\\                                      
\hline
1302+356&13 04 34.495&35 23 33.534&46.8&5.9&$-$&11.0&6.0  &97\\
        &13 04 34.494&35 23 33.538&60.5&$-$&$-$&15.0&7.0&147\\
\hline
1407+369&14 09 09.504&36 42 08.195&81.0 &1.9 &$-$ &17.0&3.0&138\\
        &14 09 09.508&36 42 08.164&192.8&76.7&42.0&8.0&1.5&141\\
        &14 09 09.508&36 42 08.152&$-$  &9.5 &4.8 &0.7&0.2&140 \\
\hline
1627+289&16 29 12.264&28 51 34.062&111.5&8.1&$-$&10.0&6.0&58\\
\hline
\end{tabular}
\end{center}

\vspace{0.5cm}
{\small
Description of the columns:
(1) source name in the IAU format;
(2) component right ascension (J2000) as measured at 1.7\,GHz;
(3) component declination (J2000) as measured at 1.7\,GHz;
(4) VLBA flux density in mJy at 1.7\,GHz from the present paper;
(5) VLBA flux density in mJy at 5\,GHz from the present paper;
(6) VLBA flux density in mJy at 8.4\,GHz from the present paper;
(7) deconvolved component major axis angular size at 1.7\,GHz obtained using
JMFIT;
(8) deconvolved component minor axis angular size at 1.7\,GHz obtained using
JMFIT;
(9) deconvolved major axis position angle at 1.7\,GHz obtained using JMFIT.
In the case the component is not visible in 1.7\,GHz map the values for the
last three columns are taken from the 5-GHz image.}
\label{table3}
\end{table*}

The projected linear size {\it D} of a radio quasar or radio galaxy can be
approximately related to the time, from the triggering of activity,
as the relationship between these variables is only weakly dependent upon
the radio luminosity. Using the model of radio source evolution from
\citet{willott99}, the age of 1045+352 was estimated to be
$\sim 10^{5}$ years \citep[see also][]{willott02, raw04}. For the
calculations we assumed: $\theta=20\degr, \beta=1.5, c_{1}=2.3,
n_{100}=3000\,{\rm e^{-}~m^{-3}}, a_{0}=100\,{\rm kpc}$ \citep[see][for
definitions]{willott99}. Both the MERLIN
and VLBA high frequency images have revealed that two cycles of activity may 
have occurred during these $\sim10^{5}$ years. The extended NE/SW emission 
is probably the remnant of the first phase of activity, which has been very
recently replaced by a new phase of activity pointing in a NW/SE
direction. It has been shown by \citet{stan05} that the extended emission 
observed for small-scale objects can be the remnants of an earlier period of 
activity in these sources. 
In the case of 1045+352, renewal of activity has been accompanied by a reorientation 
of the jet axis. 

Several processes can be used to explain a jet reorientation in AGNs.
There are strong observational and theoretical grounds for believing
that accretion disks around black holes may be twisted or warped, and this
can be caused by a number of possible physical processes. In particular, if
there is a misalignment between the axis of rotating black hole and the axis of
its rotating accretion disk, then the Lense-Thirring precession produces a warp
in the disk. This process is called the Bardeen-Peterson effect
\citep{bardeen75}. According to \citet{pringle97}, disk warping can also be induced by
internal instabilities in the accretion disk caused by radiation pressure from
the central source.

A reorientation of the jet axis may also result from a merger with another
black hole. 
\citet{merritt02} have shown that a rapid change in jet orientation can be caused
by even a minor merger because of a
spin-flip of the central active black hole arising from the coalescence of
inclined binary black holes. According to \citet{liu04},
the Bardeen-Peterson effect can also cause a realignment of a rotating SMBH
and a misaligned accretion disk, where the timescale of such a realignment
$t<10^{5}$ years. If it is assumed that the typical speed of advance of radio 
lobes of young AGNs is $\upsilon\sim$0.3$c$ \citep{ocp98,gir03,pol03}, then
distorted jets of length, $t\upsilon<$10 kpc for some CSS and GPS sources should be
observed, although the character of these disturbances is not known.   
\citet{liu04} shows that the interaction/realignment of a
binary and its accretion disk leads to the development of X-shaped sources. 
1045+352 is not a typical X-shaped source like 3C\,223.1 or 3C\,403 
\citep{dennett02,capetti02}. However, according to \citet{cohen05} the realignment 
of a rotating SMBH followed by a repositioning of the
accretion disk and jets is a plausible interpretation for misaligned radio
structures, even if they are not conspicuously X-shaped. 
 
It is likely that in young sources such as 1045+352, the gas has not
yet settled into a regular disk following a merger event and that separate 
clouds of gas and dust reaching the very central regions of the
source at different times disturb the stability of the accretion disk and
affect the jet formation. Later, these clouds could cause a renewal of activity.   
Numerical simulations of colliding galaxies show that these usually
merge completely after a few encounters in timescales up to $\sim 10^8$
years \citep{barnes96}. According to \citet{schoen00}, multiple encounters between
interacting galaxies can cause interruptions of activity and lead to the many 
types of sources that are observed in a restarted phase, such as
double-double radio galaxies. Nevertheless, it is unclear whether such encounters
can cause jet reorientation. On the other hand, the dense medium of a host 
galaxy can frustrate the jets, and their collisions with the dense
surrounding medium can cause rapid bends through large angles. In the case
of 1045+352, the VLBA images at the higher frequencies seem to show a jet 
emerging in a S/SE direction, but being bent through 
$\sim60\degr$ to 
a NE direction in the lower resolution 1.7-GHz image. The MERLIN 
lower resolution 5-GHz image might indicate that the jet has been bent again
and now emerges from the core in a NW direction.

It is difficult to find a convincing argument in favour of one of the 
above-mentioned alternatives or to rule
any of them out based upon the extensive multifrequency data on 1045+352
presented here. However, if it
is assumed that a merger is the most probable cause of the ignition and
restart of activity in radio galaxies,
this could mean that 1045+352 has undergone two merger events in a very
short period of time ($\sim 10^{5}$), which is unlikely. More probable is 
that the ignition of activity in
1045+352 has occurred during a merger
event that is, as yet, incomplete and that disturbed, misaligned radio
jets result from the realignment of a
rotating SMBH or intermittent gas injection that interrupts jet formation.
 
\subsection{Other nine sources}
Three sources from our sample (1126+293, 1407+369, 1627+289) show one- or
two-sided core-jet
structures, indicating that they are in an active phase of their evolution, although
the core-jet structure of 1126+293 is controversial. Our images indicate that the western
components are parts of the jet, which is possibly precessing or being bent
by interactions with the interstellar medium. They could, however, also be hotspots
of a radio lobe. Unfortunately, our high frequency VLBA observations are not sensitive 
enough to settle this problem. Three other sources (1056+316, 1132+374, 1425+287) have 
visible radio cores and parts of lobes or hotspots, indicating activity. 1132+374 is a
CSO object. In the case of one source,
1059+351, the VLBA observations show only a radio core, although the 5-GHz 
MERLIN image of 1059+351 also 
shows remnants of the two radio lobes of its ``S'' shaped structure visible
at the VLA resolutions \citep{mach83, mach98}. According to \citet{taylor96} and
\citet{r96}, 
``S'' symmetry is observed in many compact sources and can be explained by precession of the
central engine. 1059+351 is the largest source in our sample with a linear size
of 45 kpc based upon its largest angular size measured from 1.46-GHz VLA image \citep{mach83}.

The compact 1049+384 and 1302+356 steep spectrum sources appeared to be low-frequency 
variables (LFV) at 151\,MHz with very high ($\geq$0.99)
probabilities that their variability is real \citep{minns00}.
According to them, LFV objects are generally more compact than
other CSS sources and tend to
exhibit steeper spectra than typical CSS sources. This may be because
of rapid spectral ageing, which might be expected for frustrated sources, or
it might simply be because the sources are at very high redshifts.

\section{Conclusions}

VLBA, VLA, and MERLIN images of ten compact steep spectrum sources have been
presented. One of these sources, 1045+352, is a very radio-luminous BAL quasar, 
whose complex structure suggests restarted activity. This may have 
resulted either from a merger event or from the infall of a cloud of gas,
that had cooled in the halo of the galaxy into the core region of the
source. The asymmetric radio jets of 1045+352 and the estimated angle 
suggest that some of the emission can be boosted, although the intrinsic
asymmetries cannot be ruled out.
It has also been confirmed that the 850\,$\mu$m flux of 1045+352 can be 
severely contaminated by synchrotron emission, which may suggest less than
previously estimated values of infrared emission and dust mass. 
Most of the radio-loud BAL quasars detected
to date have very compact radio
structures similar to GPS and CSS sources which are thought to be young.
Therefore, the compact structure and young age of 1045+352 fit well to the 
evolutionary interpretation of radio-loud BAL QSOs.

According to the evolutionary model recently proposed by \citet{lipari06}, BAL 
quasars are young systems with composite outflows, and they are accompanied by absorption
clouds. The radio-loud systems may be associated with the later stages of evolution, 
when jets have removed the clouds responsible
for the generation of BALs. The effect of orientation could play a
secondary role here. The above could explain the rarity of extended radio structures 
showing BAL features \citep{gregg06}.

\begin{acknowledgements}

\item The VLBA is operated by the National Radio Astronomy Observatory
(NRAO), a facility of the National Science Foundation (NSF) operated under
cooperative agreement by Associated Universities, Inc. (AUI).

\item This research has made use of the NASA/IPAC Extragalactic Database
(NED), which is operated by the Jet Propulsion Laboratory, California
Institute of Technology, under contract with the National Aeronautics and
Space Administration.

\item Use has been made of the Sloan Digital Sky
Survey (SDSS) Archive. The SDSS is managed by the Astrophysical
Research Consortium (ARC) for the Participating Institutions: 
The University of Chicago, Fermilab, the
Insti\-tute for Advanced Study, the Japan Participation Group, The Johns
Hopkins University, Los Alamos National Labora\-tory, the
Max-Planck-Institute for Astronomy (MPIA), the Max-Planck-Institute for
Astrophysics (MPA), New Mexico State University, University of Pittsburgh,
Princeton University, the United States Naval Observatory, and the
University of Washington.

\item We thank M. Gawro\'nski for his help with the OCRA-p observations. 
The OCRA project was supported by the Polish Ministry of Science and Higher 
Education
under grant 5 P03D 024 21 and the Royal Society Paul Instrument Fund.

\item We thank P.J. Wiita for a discussion and P. Thomasson
for reading of the paper and a number of suggestions.

\item This work was supported by the Polish Ministry of Science and Higher
Education under grant 1 P03D 008 30.

\end{acknowledgements}

\end{document}